\documentclass[11pt]{article}

\usepackage{color,graphicx}
\usepackage{young}
\usepackage[vcentermath]{youngtab}
\usepackage{amsmath,amssymb,graphicx}
\usepackage{hyperref}
\definecolor{darkred}{rgb}{0.65,0.15,0}
\hypersetup{pdfborder={0 0 0},colorlinks=true,urlcolor=darkred,citecolor=blue,linkcolor=darkred,linktocpage=true}

\usepackage{cite}
\usepackage{amsmath}
\usepackage{amsfonts}
\usepackage{amssymb}
\usepackage{graphicx}%
\usepackage{amsthm}
\usepackage{mathrsfs}
\usepackage[T1]{fontenc}
\usepackage{enumerate}
\usepackage{color}
\newtheorem{theorem}{Theorem}[section]

\theoremstyle{definition}
\newtheorem{definition}[theorem]{Definition}

\def\4diml{four-dimensional}

\def\-1{^{-1}}

\newcommand{\A}{\mathscr{A}}
\newcommand{\B}{\mathscr{B}}
\newcommand{\C}{\mathscr{C}}
\newcommand{\M}{\mathscr{M}}
\newcommand{\D}{\mathscr{D}}
\newcommand{\G}{\mathscr{G}}

\makeatletter

\@addtoreset{equation}{section}
\makeatother

\begin{document}

\thispagestyle{empty}

\vspace{5mm}

\begin{center}
{\LARGE \bf More on (gauged) WZW models over low-dimensional  \\[2mm] Lie supergroups and their integrable deformations}

\vspace{15mm}

\normalsize
{\large  Ali Eghbali\footnote{Corresponding author: eghbali978@gmail.com}, Meysam Hosseinpour-Sadid\footnote{meysam.hs.az@gmail.com},
Adel Rezaei-Aghdam\footnote{rezaei-a@azaruniv.ac.ir}}\\

\vspace{2mm}
{\small \em Department of Physics, Faculty of Basic Sciences,\\
Azarbaijan Shahid Madani University, 53714-161, Tabriz, Iran}\\
\vspace{10mm}
\vspace{6mm}

\begin{tabular}{p{12cm}}
{\small

In superdimension $(2|2)$ there are only three non-Abelian Lie
superalgebras admitting non-degenerate ad-invariant supersymmetric metric, the well-known
Lie superalgebra  $gl(1|1)$, and two more, $({\C}^3 + \A)$ and $({\C}_0^5 +{\A})$.
After a brief review of the construction of the Wess-Zumino-Witten (WZW) models based on the $GL(1|1)$ and $(C^3 + A)$ Lie supergroups,
we proceed to construct the WZW model on the $({C}_0^5 +{A})$ Lie supergroup. Unfortunately, this model does not include
the super Poisson-Lie symmetry.
In the following, three new exact conformal field theories of the WZW type are constructed by gauging an anomaly-free subgroup
SO(2) of the Lie supergroups mentioned above.
The most interesting indication of this work is that the gauged WZW model on the supercoset $(C^3 + A)/$SO(2)
has super Poisson-Lie symmetry; most importantly, its dual model is conformally invariant at the one-loop order,
and this is presented here for the first time.
Finally, in order to study the Yang-Baxter (YB) deformations of the $({C}_0^5 +{A})$ WZW model
we obtain the inequivalent solutions of the
(modified) graded classical Yang-Baxter equation ((m)GCYBE) for the $({\C}_0^5 +{\A})$ Lie superalgebra.
Then, we classify all possible YB deformations for the $({C}_0^5 +{A})$
and settle also the issue of an one-loop conformality of the deformed backgrounds.
The classification results are important, in particular in the Lie supergroup case they are rare, much hard technical work was needed to obtain them.
}
\end{tabular}
\vspace{-1mm}
\end{center}
{\small
{$~~~~~~~~~$\bf Keywords:}  Lie supergroup, $\sigma$-model, String duality, Wess-Zumino-Witten model,\\
$~~~~~~~~~~~~~~~~~~~~~~~~~~~$Yang-Baxter deformation}

\setcounter{page}{1}
\newpage
\tableofcontents

 \vspace{5mm}

\section{Introduction}
\label{Sec.1}
Over the decades of study in string theory, it has been shown that
finding exact string backgrounds is a very important problem in this theory.
These determine the short-range structure of spacetime and provide
information about the gravitational interactions generated by the strings.
A class of exact solutions to string theory, which are roughly distinct from each other,
can be listed as follows: flat space with linear dilaton \cite{ref.1},
plane wave \cite{ref.2}, as well as $N = 4$ supersymmetric string backgrounds \cite{ref.3}, WZW models
\cite{ref.4} and gauged WZW models \cite{{ref.5},{ref.6},{ref.7},{ref.8},{ref.9}}.
Indeed, the last two cases are very important in understanding string theory
since they provide exact solutions to it and the current algebra description makes the
theories solvable.
We will examine here solutions in the last two classes writing down all ungauged and gauged WZW models based on Lie supergroups of the types $(2|2)$ and $(1|2)$.
Indecomposable Lie superalgebras with superdimension up to four were first classified by Backhouse in \cite{B}.
Among Lie superalgebras of type $(2|2)$, there are three non-Abelian
superalgebras which have non-degenerate ad-invariant supersymmetric metric, the well-known
Lie superalgebra  $gl(1|1)$, and two more, $({\C}^3 + \A)$ and $({\C}_0^5 +{\A})$.

Gauged WZW models are a natural framework for giving a Lagrangian realization of
coset models. A wide class of conformal field theories can be classified as $G/H$ coset models using the Goddard, Kent and Olive (GKO)
algebraic construction \cite{GKO}.
The GKO construction relies on the relationship between affine Kac-Moody algebras and Virasoro algebras.
It was originally proposed by Schnitzer, Karabali, and others that the vector gauged WZW theory \cite{H.J.Schnitzer,ref.9} is
the GKO coset construction $G/H$ in conformal field theory.
However, this proposal was not proved except when $H$
is an Abelian gauge group. Then, in Ref. \cite{wei.Chung},  the equivalence of gauged WZW theory and coset model
in conformal field theory was demonstrated for both vector and chiral.

Conformal $\sigma$-models and WZW models on supercoset spaces provide important examples of logarithmic conformal field theories \cite{{rosansky},{Gaberdiel}},
because of their applications in string theory and condensed matter theory.
Investigations of algebraic and mostly chiral aspects of WZW models based on Lie supergroups reach back more than thirty years \cite{saleur1}.
We note that the special properties of Lie supergroups allow for constructions which are not possible for ordinary Lie groups.
For example, there exist several families of coset conformal field theories which are obtained by gauging a
one-sided action of some subgroup rather than the usual adjoint \cite{Metsaev}.
The same class of corresponding $\sigma$-models admit a kind of marginal deformations that are not of current-current type \cite{Berkovits.Vafa}.
One could think of including the case of ${AdS}_4 \times CP^3$ as another explicit interesting example of an integrable string background.
In \cite{Sergey.Frolov}, it has been argued that the Green-Schwarz action for type IIA
string theory on ${AdS}_4 \times CP^3$ with $\kappa$-symmetry partially fixed can be understood as a
coset $\sigma$-model on the space supplied with a proper WZ term (see, also, \cite{Stefanski}).
In addition to these one can find a comprehensive discussion of WZW models on target supergroups of type I in \cite{Volker.Schomerus}.
The WZW model on the $GL(1|1)$ was first studied in order to present a number of
interesting features including non-compactness, non-simplicity, $1/k^2$ quantum corrections and logarithms in
the current blocks \cite{rosansky} (see, also, \cite{{Schomerus1},{Schomerus2}}).
In \cite{ER7}, the $GL(1|1)$ WZW model was constructed in order to study the super Poisson-Lie symmetry \cite{{ER2},{ER5}} (see, also, \cite{falk.Hassler})
of the model\footnote{
During the last years, Poisson-Lie T-duality proposed by Klimcik and Severa \cite{{Klim1},{Klim2}} has been the subject of an increasing
number of papers, especially in the WZW models.
As the first example of the Poisson-Lie symmetry in the WZW models, it was shown that \cite{Klimcik4}
the Poisson-Lie T-duality relates the SL$(2,\mathbb{R})$ WZW model to a $\sigma$-model defined on the SL$(2,\mathbb{R})$ group space.
Another interesting example is the study of this symmetry in WZW model on the Heisenberg Lie group \cite{eghbali11} (see, also, \cite{{Lledo},{EMR13},{Exact},{sakatani1},{sakatani.Satoh},{ali.Fortsch}}).
Of course, before these, (non-)Abelian T-duality symmetry had been investigated in the WZW model \cite{Lozano1}.}.
Following this, the WZW model on the $({C}^3 + A)$ Lie supergroup was constructed in Ref. \cite{ER8}, and it was then shown that this model has also super Poisson-Lie symmetry (see, also, \cite{ER.super1}).
But so far, the $({C}_0^5 +{A})$ WZW model has not been studied from this point of view. In this paper we proceed to construct
WZW model based on the $({C}_0^5 +{A})$ Lie supergroup for two purposes:
investigating super Poisson-Lie symmetry and YB deformation of the model.
It is worth mentioning that unlike the bosonic Lie algebras, there is no non-degenerate metric for non-Abelian Lie superalgebras with dimension up
to three. This means that the WZW model does not exist on three-dimensional Lie supergroups,
and we can only rely on other conformally invariant models such as the gauged WZW models.
In the present work, three exact conformal field theories describing new supergeometries of type $(1|2)$ are found as the gauged WZW models
on the supercosets $GL(1|1)/$SO(2),  $(C^3 + A)/$SO(2) and $({C}_0^5 +{A})/$SO(2).
Then, we show that the gauged WZW model on the supercoset $(C^3 + A)/$SO(2) has super Poisson-Lie symmetry,
in such a way that we find its dual pair.

In the bosonic case, one of the most famous WZW models was based on the group $E^2_c$, a central extension
of the two-dimensional Euclidean group such that the corresponding $\sigma$-model describes string
propagation on a four-dimensional gravitational plane wave background \cite{ref.10.witten}.
This construction was subsequently extended to other non-semisimple Lie groups \cite{Sfetsos}, in such a way that the
WZW model on the Heisenberg Lie group with arbitrary dimension was introduced by Kehagias and Meessen \cite{Kehagias}
(see, also, \cite{kehagias1,Chakraborty}).
In addition, Witten had shown that \cite{ref.6} that a simple gauged WZW model \cite{{ref.5},{ref.9}} yields a
two-dimensional black hole so that the result of his work raises the possibility of using similar constructions to find
exact conformal field theories for higher dimensional black holes or extended
black holes. As in this regard, it was shown that \cite{ref.7} a simple extension of Witten's construction yields
three-dimensional charged black strings.

In the last few years, the integrable deformations of $\sigma$-models have received considerable attention.
Klimcick introduced YB $\sigma$-models \cite{{Klimcik1},{Klimcik2}} as a class of integrable deformations of $\sigma$-models.
The initial input for construction of this type of the models
is classical $r$-matrix solving the CYBE. According to the researches done in this direction, there is a variety of $r$-matrices,
giving various deformations of integrable $\sigma$-models.
One of the most interesting and first YB deformations is related to the $AdS_5 \times S^5$ superstring which was built based on
the standard inhomogeneous solution of the CYBE \cite{{Delduc.JHEP2013},{Delduc2},{Delduc3},{Arutyunov1}}.
Later, this type of deformations was rapidly developed by many researchers in the superstring backgrounds \cite{Kawaguchi1}.
The generalization to YB $\sigma$-models with WZW term has also
carried out in \cite{{Delduc4},{Yoshida.NPB},{Klimcik.PLB},{Demulder.JHEP},{Klimcik.LMP},{B.Hoare},{N.Mohammedi.PRD}} (see, also, \cite{{Epr1},{Epr3},{Daniele1}}).
In Ref. \cite{Demulder.JHEP}, by introducing the YB WZW model together with its integrability properties,
it has been given an explicit derivation of the one-loop beta-functions of the YB WZ model
in the case of arbitrary groups. Moreover, there the YB WZ action within
the framework of the ${\cal E}$-model and its corresponding Poisson-Lie T-dual model have been formulated.
Lately, YB deformations of the WZW models besed on the $GL(1|1)$ and $(C^3 + A)$ Lie supergroups have been carried out by using
the super skew-symmetric classical r-matrices satisfying (m)GCYBE \cite{Epr2}. Most interesting, it has been shown that the resulting deformed geometries,
in addition to being integrable, remain conformally invariant up to the one-loop order.
The general procedure applied in \cite{Epr2} is a straightforward generalization of
the well-known prescription of Delduc, Magro and Vicedo \cite{Delduc4}.
As a spin off from this progress, it would be interesting to study YB deformation of the $({C}_0^5 +{A})$  WZW model.
The motivation behind this study stems from the recent spate of interest in
YB deformations of the WZW models constructed in \cite{Epr2}.

There are quite a few novel results in this paper and so now is an opportune moment to summarize them:
\begin{itemize}

\item After a review of the construction of the WZW models on the $GL(1|1)$ and $({C}^3 +{A})$
Lie supergroups and investigating their super Poisson-Lie symmetry,
we construct the WZW model based on the $({C}_0^5 +{A})$ Lie supergroup to obtain a new exact conformal field theory
of the type $(2|2)$.

\item Three new $(1|2)$-dimensional exact conformal field theories of the WZW type are found by gauging an anomaly-free subgroup
H=SO(2) of the Lie supergroups mentioned above.
The most interesting indication of this work is that the gauged WZW model on the supercoset $(C^3 + A)/$SO(2)
has the super Poisson-Lie symmetry. Using the super Poisson-Lie T-duality transformations we find the corresponding dual pair, and show that
it is also conformally invariant at the one-loop order.

\item We solve the (m)GCYBE for the $({\C}_0^5 +{\A})$ Lie superalgebra to obtain the corresponding super skew-symmetric $r$-matrices.
Employing the automorphism supergroup of the $({\C}_0^5 +{\A})$, we show that the $r$-matrices are split into five inequivalent classes.
Then, YB deformations of the $({C}_0^5 +{A})$ WZW model are specified by the r-matrices satisfying (m)GCYBE.
After checking the conformal invariance of the deformed models up to one-loop order, it is concluded that
the $({C}_0^5 +{A})$ WZW model is a conformal theory within the classes of the YB deformations preserving the conformal invariance.

\end{itemize}

The paper is organized as follows. In section \ref{Sec.2}, we review the
construction of WZW models based on the $(2|2)$-dimensional Lie supergroups,
and in particular we discuss their super Poisson-Lie symmetry.
Sections \ref{Sec.3} and \ref{Sec.4} contain two important results of the work: in section \ref{Sec.3},
three new exact conformal field theories of the WZW type are obtained by gauging an anomaly-free subgroup
SO(2) of the $GL(1|1)$, $({C}^3 +{A})$ and $({C}_0^5 +{A})$ Lie supergroups.
The study of super Poisson-Lie symmetry in the supercoset $\sigma$-model on the $(C^3 + A)/$SO(2) is devoted to section \ref{Sec.4}.
Also, the corresponding dual pair is built at the end of this section.
We briefly review the formulation of the YB deformation of WZW model over Lie supergroups in section \ref{Sec.5}.
In section \ref{Sec.6}, by solving the (m)GCYBE for the $({\C}_0^5 +{\A})$ Lie superalgebra we obtain the
corresponding $r$-matrices, and show that they are split into five inequivalent classes.
The YB deformations of the $({C}_0^5 + {A})$ WZW model are also given at the end of section \ref{Sec.6}.
We make concluding remarks and discuss our results in section \ref{Sec.7}.

\section{WZW models on Lie supergroups of the type $(2|2)$ and their super Poisson-Lie symmetry}
\label{Sec.2}

The subject of this section is a brief overview of the WZW models with target spaces Lie supergroups $GL(1|1)$ and $(C^3 + A)$.
In particular, we construct WZW model based on the $({C}_0^5 +{A})$ Lie supergroup to obtain a new exact conformal field theory
of the type $(2|2)$, and then focus on the investigation of its super Poisson-Lie symmetry.
Before proceeding to doing these,
let us recall some of the properties and definitions related to $\mathbb{Z}_2$-graded vector
space and Lie superalgebras. First of all, we define a supervector space $V$.
\vspace{-2mm}
\begin{definition}\label{def1.A}
A \emph{supervector space} ${\mathbb V}$ is a ${\mathbb{Z}}_{2}$-graded vector space, a vector space as ${\mathbb V}= {\mathbb V}_{_0} \oplus {\mathbb V}_{_1}$ over an arbitrary field $\mathbb{F}$ with a given decomposition of subspaces ${\mathbb V}_{_0}$ and ${\mathbb V}_{_1}$ with grades $0$  and $1$, respectively \cite{{Kac},{N.A}}.
\end{definition}
\vspace{-2mm}
The parity of a non-zero homogeneous element, denoted by $|x|$, is $0$ (even) or $1$ (odd) according to
whether it is in ${\mathbb V}_{0}$ or ${\mathbb V}_{1}$, namely,
$|x|=0$ for any $x \in {\mathbb V}_{_0}$, while for any $x \in {\mathbb V}_{_1}$ we have $|x|=1$.
The even elements are sometimes called bosonic, and the odd elements fermionic. From now on,
we use $B$ and $F$ instead of $0$ and $1$, respectively.
\begin{definition}\label{def2.A}
A \emph{Lie superalgebra} ${\G}$ is a $\mathbb{Z}_2$-graded vector space, thus admitting the decomposition
${\G} ={\G}_{{_B}} \oplus {\G}_{_F}$, equipped with a bilinear
superbracket structure $[. , .]: {\G} \otimes {\G} \rightarrow {\G}$ satisfying the requirements of super anti-symmetry
and super Jacobi identity \cite{{Kac},{N.A}}.
\end{definition}
\vspace{-2mm}
If ${\G}$ is finite-dimensional and the dimensions of ${\G}_{_B}$ and $ {\G}_{_F}$
are $m=\#B$ and  $n=\#F$, respectively, then ${\G}$ is said to have superdimension $(m|n)$.
We shall identify grading indices by the same indices in the power of $(-1)$, i.e., we use $(-1)^x$ instead of  $(-1)^{|x|}$, where
$(-1)^x$ equals 1 or -1 if the Lie sub-superalgebra element is even or odd, respectively\footnote{
This notation was first used by Dewitt in \cite{D}. Throughout this paper we work with Dewitt's notation.}.

In the following we shall define the WZW model on a two-dimensional worldsheet $\Sigma$ with a
Lie supergroup $G$ as target space.
The field content of the theory is a supergroup-valued field $g: \Sigma \rightarrow G$, which is the
embedding map of the source space $\Sigma$ into the target supergroup $G$.
The model is a non-linear $\sigma$-model
whose action is a functional of a field $g$.
Let $\{T_{_a}\},~a= 1, \cdots, dim\hspace{0.4mm}G$ be a basis in the Lie superalgebra $\G$ of $G$,
with $V_{_a}$ denoting the corresponding left-invariant supervector fields.
The left-invariant dual super one-forms $L^a$ are defined in the usual way: $(-1)^b L^a V_{_b} ={\delta}^a_{~b}$.
The Maurer-Cartan left-invariant super one-form on $G$ given by $L=(-1)^a L^a T_{_a}=(-1)^a (g^{-1}dg)^a T_{_a}$, shall be needed
in order to define the WZW model on the group supermanifold. They satisfy the super Maurer-Cartan equation
$d  {L}^{^a}= -{1}/{2} ~(-1)^{^{bc}} {f^a}_{bc} ~ {L}^{^b} \wedge {L}^{^c}$.
Any invariant metric $\big<.~,~.\big> $ on $\G$ can be induced a bi-invariant metric on $G$ defined by $\big<V_{_a} , V_{_b}\big>$.
We define a non-degenerate ad-invariant supersymmetric metric on Lie superalgebra ${\G}$ as $\Omega_{_{ab}}=\big<T_{_a} , T_{_b}\big>$
which is needed to define the WZW model.
One may use the ad-invariant inner product on $\G$ to obtain \cite{ER7}
\begin{eqnarray}\label{2.1}
f^d_{~ab} \;\Omega_{dc}+(-1)^{bc} f^d_{~ac} \;\Omega_{db}\;=\;0,
\end{eqnarray}
where $f^c_{~ab}$ are the structure constants of Lie superalgebra ${\G}$.
The action of WZW model is thus given by\footnote{Note that for WZW models on bosonic Lie groups one usually introduces an integer valued constant,
the level $k$, appearing as a prefactor of the Killing form. For supergroups the Killing form might vanish.
For simple supergroups the level has a well-defined meaning in the sense of multiplying a standard non-degenerate invariant form. There may also be quantization conditions linked to the topology of the supergroup. Here we deal with non-semisimple examples of supergroups.
Moreover, we would like to include models whose metric renormalizes non-multiplicatively. Under these circumstances it is not particularly convenient to display the level explicitly and we assume instead that all possible parameters are contained in the metric.
}
\begin{eqnarray}\label{2.2}
S_{_{_{\hspace{0mm}WZW}}}(g)=\frac{1}{2}\int_{_\Sigma}d\sigma^+ d\sigma^- \big< g^{-1} \partial_{+}g , \; g^{-1}
\partial_{-}g \big>   +\frac{1}{12}\
\int_{_{B_{_3}}}d^{3}\sigma~\big< g^{-1}dg\; \hat{,}
\;[g^{-1}dg \;\hat{,}\;g^{-1}dg]
 {\big>},~~~
\end{eqnarray}
where $\sigma^\alpha =(\sigma^+ , \sigma^-)$ are the standard lightcone variables such that their relationship with
the worldsheet coordinates $(\tau , \sigma)$ is given by $\sigma^{\pm} = {(\tau\pm\sigma)}/\sqrt{2}$.
Here, the first term contains the dynamics while the second one is the so-called WZ term which features an integration of a suitable
extension of $g$ over a three-manifold ${B_{_3}}$ parameterized
by $(\tau, \sigma, \xi)$ whose boundary is the worldsheet, where
the extra direction is labeled by $\xi$.
For future convenience the WZW action \eqref{2.2} can be written explicitly in terms of the components of left-invariant super one-forms
$L^{\hspace{-0.5mm}a}_{\pm}$'s, giving us \cite{ER7}
\begin{eqnarray}\label{2.3}			
S_{_{_{\hspace{0mm}WZW}}}(g)=\frac{1}{2}\int_{_\Sigma}d\sigma^+ d\sigma^-  ~ (-1)^a L_{+}^{a} \Omega_{_{ab}} L_{-}^{b}  +\frac{1}{12}\
\int_{_{B_{3}}}d^{3}\sigma  (-1)^{a+bc}  \varepsilon^{\alpha\beta\gamma} L_{\alpha}^{a} \Omega_{ad}~ f^d_{~bc} ~ L_{\beta}^{b} L_{\gamma}^{c},~~
\end{eqnarray}
where $ \varepsilon^{\alpha\beta\gamma}$ stands for the Levi-Civita symbol.
One may regard the WZW model \eqref{2.3} as a $\sigma$-model in the following form
\begin{eqnarray}
S&=& \frac{1}{2}\int_{_{\Sigma}}\!d\sigma^+  d\sigma^- ~ (-1)^{\mu} \partial_{_+} x^{^\mu} {G}_{_{\mu \nu }} \partial_{_-} x^{^\nu}
+ \frac{1}{12}\int_{_{B_{3}}}\!d^3\sigma~  (-1)^{\mu} \varepsilon^{^{\gamma \beta \alpha}}
\partial_{_\gamma} x^{^\mu}  {H}_{_{\mu \nu \rho}} \partial_{_\alpha} x^{^\rho} \partial_{_\beta} x^{^\nu}\nonumber\\
&=& \frac{1}{2}\int_{_{\Sigma}}\!d\sigma^+  d\sigma^- ~ (-1)^{\mu} \partial_{_+} x^{^\mu} ({G}_{_{\mu \nu }}+{B}_{_{\mu \nu }}) \partial_{_-} x^{^\nu},\label{2.4}
\end{eqnarray}
where ${G}_{_{\mu \nu }}$ and ${B}_{_{\mu \nu }}$ are the respective supersymmetric metric and
super anti-symmetric two-form field ($B$-field) on supergroup $G$ with the coordinates
$x^{^\mu},~\mu =1, \cdots, dim\hspace{0.4mm} G$\footnote{The functions $x^{^\mu}$ include the
bosonic coordinates $x^i$ and the fermionic ones $\theta^{^\alpha}$,
and the labels $\mu, \nu$ run over $i =0,\cdots, d_{_B}- 1$ and $\alpha = 1,\cdots, d_{_F}$, where $(d_{_B}|d_{_F})$ denotes
the superdimension of $G$.}. Moreover,
$H_{_{\mu\nu\rho}}$ is the field strength of the $B$-field which is defined as follows:
\begin{eqnarray}\label{2.5}
H_{_{\mu\nu\rho}}=(-1)^{^{\mu}}\;\frac{\overrightarrow{\partial}}{\partial
x^{^{\mu}}} B_{_{\nu\rho}}+(-1)^{^{\nu+\mu(\nu+\rho)}}\;\frac{\overrightarrow{\partial}}{\partial
x^{^{\nu}}} B_{_{\rho\mu}}+(-1)^{^{\rho(1+\mu+\nu)}}\;\frac{\overrightarrow{\partial}}{\partial
x^{^{\rho}}} B_{_{\mu\nu}}.
\end{eqnarray}
By comparing \eqref{2.3} and \eqref{2.4},  one then finds that ${G}_{_{\mu\nu}}$ and $H_{_{\mu\nu\rho}}$  can be written in the following forms\footnote{Here the superscript ``st'' in $L^{^{st}}$ stands for the supertranspose \cite{D}.}
\begin{eqnarray}
{G}_{_{\mu\nu}} &=& (-1)^a~ {L_{_\mu}}^{^a}~\Omega_{_{ab}}~{({L^{st})}^{^b}}_{_\nu},\label{2.5.1}\\
H_{_{\mu\nu\rho}} &=& -(-1)^{^{a+b\nu}} {L_{_\mu}}^{^a}~  \Omega_{_{ad}}  {f^d}_{bc} ~ {({L^{st})}^{^c}}_{_\nu}~ {({L^{st})}^{^b}}_{_\rho}.\label{2.5.2}
\end{eqnarray}
$\bullet$~{\it The WZW model on the $GL(1|1)$ Lie supergroup.}\\
The $gl(1|1)$ Lie superalgebra\footnote{In Backhouse's classification \cite{B},
the $gl(1|1)$ Lie superalgebra has been labeled  by $({\C}_{-1}^2+{\A})$.} is spanned by the set of generators $\{T_{_1}, T_{_2}; T_{_3}, T_{_4}\}$
with gradings, grade$(T_{_1})$ = grade$(T_{_2})$ = 0 and grade$(T_{_3})$ = grade$(T_{_4})$ = 1,\footnote{From now on we denote the bosonic generators
by $(T_{_1}, T_{_2})$ and fermionic ones by $(T_{_3}, T_{_4})$.}
which fulfill the following (anti)commutation relations
\begin{eqnarray}\label{2.6}
[T_{_1} , T_{_3}]=T_{_3},~~~~~~~[T_{_1} , T_{_4}]=-T_{_4},~~~~~~~\{T_{_3} , T_{_4}\}=T_{_2}.
\end{eqnarray}
It can be easily shown that the most general ad-invariant bilinear form on the $gl(1|1)$ is
\begin{eqnarray}\label{2.7}
\Omega_{ab}=\left( \begin{tabular}{cccc}
                $\beta$ & 1  &  0        & 0\\
                1     & 0  &  0        & 0\\
                0     & 0  &  0        & 1\\
                0     & 0  & -1 & 0\\
                 \end{tabular} \right),
\end{eqnarray}
for some constant $\beta$.
In order to write \eqref{2.3} explicitly we need to find the $L^{\hspace{-0.5mm}a}_{\pm}$'s.
To this purpose we use the following parametrization of the Lie supergroup:
\begin{eqnarray}
g = e^{\chi T_4}~e^{y T_1}~e^{x T_2}~e^{\psi T_3}, \label{2.8}
\end{eqnarray}
where $x(\tau , \sigma)$ and $y(\tau , \sigma)$ are the bosonic fields
while $\psi(\tau , \sigma)$ and $\chi(\tau , \sigma)$ stand for the fermionic fields.
By using \eqref{2.6} and \eqref{2.8} we find
\begin{eqnarray}
g^{-1} \partial_{\pm} g &=& (-1)^a  L^{\hspace{-0.5mm}a}_{\pm}  T_{_a} \nonumber\\
&=&\partial_{\pm} y~ T_{_1} +   (\partial_{\pm} x -  \partial_{\pm} \chi ~\psi e^y) T_{_2}+
(\partial_{\pm} \psi +  \partial_{\pm}y ~\psi)T_{_3} +\partial_{\pm} \chi~e^y T_{_4}, \label{2.9}
\end{eqnarray}
from which we can read off the $L^{\hspace{-0.5mm}a}_{\pm}$'s and thus the Lagrangian looks like \cite{ER7}
\begin{eqnarray}\label{2.10}
S_{_{_{\hspace{0mm}WZW}}}(g) = \frac{1}{2} \int d \sigma^+ d \sigma^-\Big[\beta\partial_{_{+}} y \partial_{_{-}} y+\partial_{_{+}} y \partial_{_{-}} x+ \partial_{_{+}} x \partial_{_{-}} y
-2 e^{y} \partial_{_{+}} \psi \partial_{_{-}} \chi\Big].~~
\end{eqnarray}

The next step is that to discuss the super Poisson-Lie symmetry of the model.
Following \cite{ER2}, one says that the background ${{\cal E}_{_{\mu \nu}}}={G}_{_{\mu \nu }}+{B}_{_{\mu \nu }}$
of the action \eqref{2.4} has the super Poisson-Lie symmetry if
\begin{eqnarray}\label{2.11}
{\cal L}_{_{V_{_a}}}{\cal E}_{_{{\mu \nu}}}
=(-1)^{a + \lambda+a \mu+c \rho}~ {{\tilde f}^{bc}}_{\; \; \;a}~{\cal E}_{_{{\mu \rho}}}
{V_{_c}}^{^{\rho}}~{V_{_b}}^{^{\lambda}}~{\cal E}_{_{{\lambda \nu}}},
\end{eqnarray}
where ${\cal L}_{_{V_{_a}}}$ stands for the Lie derivative corresponding to the left-invariant supervector
fields ${V_{_a}}$ satisfying $[V_{_a} , V_{_b}]={f^c}_{ab} ~V_c$, and
$\tilde f^{ab}_{~~c}$ are the structure constants of $\tilde \G$, the Lie superalgebra dual to $\G$
\footnote{Note that the Lie superalgebras $\G$ and $\tilde \G$ form a Lie superbialgebra which is denoted by
$(\G, \tilde \G)$ \cite{N.A} (see, also, \cite{ER1}).}, whose dimension is, however,
equal to that of $\G$.
It's worth noting that the integrability condition on the Lie derivative,
$[{\cal L}_{_{V_{_a}}} , {\cal L}_{_{V_{_b}}}]= {\cal L}_{_{[V_{_a} , V_{_b}]}}$,
then implies the mixed super Jacobi identities \cite{ER2} showing that this construction leads naturally to
the Drinfeld superdouble \cite{ER4}.

In order to investigate the super PL symmetry of the WZW  model \eqref{2.10} one has to employ equation \eqref{2.11}. First,
we need to find the left-invariant supervector fields corresponding to left-invariant super one-forms given by \eqref{2.9}.
Based on this, in Ref. \cite{ER7} it has been shown that relation \eqref{2.11} holds for the WZW background of action \eqref{2.10}
if the dual pair to the $gl(1|1)$ is the ${{\B} \oplus {\A } \oplus {\A}_{1,1}}$ \cite{ER6} whose only non-zero commutation relation is
$[{\tilde T}^{^2} , {\tilde T}^{^3}]={\tilde T}^{^3}$. In fact, it can be said that
the super Poisson-Lie duality relates the
$GL(1|1)$ WZW model to a $\sigma$-model defined on the $GL(1|1)$ Lie
supergroup when the dual Lie supergroup is ${{B} \oplus {A } \oplus {A}_{1,1}}$, in such a way
the pair $\big(gl(1|1) , {{\B} \oplus {\A } \oplus {\A}_{1,1}}\big)$ as a Lie superbialgebra satisfies mixed super Jacobi identities \cite{ER6}.
\\\\
$\bullet$~{\it The WZW model on the $(C^3 + A)$ Lie supergroup}.\\
The $({\C}^3 + {\A})$ Lie superalgebra possesses four generators $\{T_{_1}, T_{_2}; T_{_3}, T_{_4}\}$
so that they obey the following set of non-zero (anti)commutation relations \cite{B}
\begin{eqnarray}\label{2.12}
[T_{_1} , T_{_4}]=T_{_3},~~~~~~~~~~\{T_{_4} , T_{_4}\}=T_{_2}.
\end{eqnarray}
The parametrization of a general element of $(C^3 + A)$ Lie supergroup we choose as in \eqref{2.8}. Then, one gets
\begin{eqnarray}\label{2.13}
g^{-1} \partial_{_{\pm}} g = \partial_{_{\pm}} y ~T_{_{1}} + (\partial_{_{\pm}} x-\partial_{_{\pm}} \chi~ \frac{\chi}{2})~T_{_{2}} +
(\partial_{_{\pm}} \psi-\partial_{_{\pm}} \chi~ y)~T_{_{3}}+\partial_{_{\pm}} \chi ~T_{_{4}}.
\end{eqnarray}
Using \eqref{2.12}, \eqref{2.13} and the fact that ad-invariant metric on the $({\cal C}^3 +{\cal A})$ is as in
\eqref{2.7}, one can compute the action of WZW model on the $(C^3 +A)$ Lie supergroup, giving us \cite{ER8}
\begin{eqnarray}\label{2.14}
S_{_{_{\hspace{0mm}WZW}}}(g) &=& \frac{1}{2} \int d \sigma^+ d \sigma^-\Big[\beta\partial_{_{+}} y \partial_{_{-}} y+\partial_{_{+}} y \partial_{_{-}} x+ \partial_{_{+}} x \partial_{_{-}} y
+\partial_{_{+}} y  \chi \partial_{_{-}} \chi\nonumber\\
&&~~~~~~~~~~~~~~~~~~~~~~~~~~~~~~~~~~~~~~~~~~~~~-\partial_{_{+}} \psi \partial_{_{-}} \chi+\partial_{_{+}} \chi \partial_{_{-}} \psi\Big].
\end{eqnarray}
In \cite{ER8}, it has been shown that the above model has the super Poisson-Lie symmetry. This means that the background
of the model satisfies the condition \eqref{2.11} with the dual Lie superalgebra ${\C}^3 \oplus {\A}_{1,1}$ \cite{ER14}
which is defined by the commutation relation $[{\tilde T}^{^2} , {\tilde T}^{^3}]=-\frac{1}{2}{\tilde T}^{^4}$.
\\\\
$\bullet$~{\it The WZW model on the $(C_0^5 + A)$ Lie supergroup}.\\
As mentioned in the Introduction section, the WZW model on the $(C_0^5 + A)$ has not been studied, until now.
Before proceeding to doing, let us introduce the $({\C}_0^5 + {\A})$ Lie superalgebra. It is a non-trivial
Lie superalgebra of the type $(2|2)$ which is defined by the following non-zero Lie superbrackets \cite{B}:
\begin{eqnarray}\label{2.17}
[T_{_1} , T_{_3}]=-T_{_4},~~~~~~~[T_{_1} , T_{_4}]=T_{_3},~~~~~~~\{T_{_3} , T_{_3}\}=T_{_2},~~~~~~~\{T_{_4} , T_{_4}\}=T_{_2}.
\end{eqnarray}
In order to calculate the left-invariant super one-forms on the $(C_0^5 + A)$ we parametrize an element of the supergroup
with the coordinates $(y, x; \psi,  \chi)$ so that its elements can be written as
\begin{eqnarray}
g = e^{\psi T_3}~e^{\chi T_4}~e^{y T_1}~e^{x T_2}. \label{2.18}
\end{eqnarray}
One then calculates
\begin{eqnarray}\label{2.19}
g^{-1} \partial_{_{\pm}} g &=& \partial_{_{\pm}} y ~T_{_{1}} + \big(\partial_{_{\pm}} x - \partial_{_{\pm}} \psi \frac{\psi}{2} - \partial_{_{\pm}} \chi \frac{\chi}{2}\big) T_{_{2}} \nonumber\\
&&~~+\big(\partial_{_{\pm}} \psi~\cos y - \partial_{_{\pm}} \chi~ \sin y\big) T_{_{3}}+ \big(\partial_{_{\pm}} \psi ~\sin y + \partial_{_{\pm}} \chi ~\cos y\big) T_{_{4}},
\end{eqnarray}
for which the $L^{\hspace{-0.5mm}a}_{\alpha}$'s are obtained to be of the following form
\begin{eqnarray}\label{2.20}
L^{\hspace{-0.5mm}1}_{\pm}&=&\partial_{_{\pm}} y,\nonumber\\
L^{\hspace{-0.5mm}2}_{\pm}&=&\partial_{_{\pm}} x - \partial_{_{\pm}} \psi \frac{\psi}{2} - \partial_{_{\pm}} \chi \frac{\chi}{2},\nonumber\\
L^{\hspace{-0.5mm}3}_{\pm}&=& - \partial_{_{\pm}} \psi~\cos y + \partial_{_{\pm}} \chi~ \sin y,\nonumber\\
L^{\hspace{-0.5mm}4}_{\pm}&=& -\partial_{_{\pm}} \psi ~\sin y - \partial_{_{\pm}} \chi ~\cos y.
\end{eqnarray}
Analogously, the ad-invariant metric on the $({\C}_0^5 + {\A})$ is as in \eqref{2.7}.
Using these and some algebraic calculations, the WZW action on the $({C}_0^5 + {A})$ Lie supergroup are worked out
\begin{eqnarray}\label{2.21}
S_{_{_{\hspace{0mm}WZW}}}(g) &=& \frac{1}{2} \int d \sigma^+ d \sigma^-\Big[\beta\partial_{_{+}} y \partial_{_{-}} y+\partial_{_{+}} y \partial_{_{-}} x+ \partial_{_{+}} x \partial_{_{-}} y
+\partial_{_{+}} y  \psi  \partial_{_{-}} \psi\nonumber\\
&&~~~~~~~~~~~~~~~~~~~~~~~~~+\partial_{_{+}} y  \chi  \partial_{_{-}} \chi
 -\partial_{_{+}} \psi ~ \partial_{_{-}} \chi +\partial_{_{+}} \chi ~ \partial_{_{-}} \psi\Big].
\end{eqnarray}
By regarding this action as a $\sigma$-model action of the form
\eqref{2.4}, we can read off the line element and $B$-field as follows:
\begin{eqnarray}
ds^2 &=& (-1)^{\mu \nu} ~G_{_{\mu \nu}} dx^\mu~dx^\nu = \beta dy^2+ 2dydx + \psi ~ dy d\psi + \chi dy d\chi -2 d\psi d\chi,~~~\label{2.22}\\
B &=& \frac{ (-1)^{\mu \nu}}{2} ~B_{_{\mu \nu}} dx^\mu  \wedge dx^\nu =\frac{\psi}{2} ~dy \wedge d\psi + \frac{\chi}{2} ~dy \wedge d\chi.\label{2.23}
\end{eqnarray}
The action \eqref{2.21} as a WZW model should be conformally invariant.
To check this, one first looks at the one-loop beta function equations \cite{ER5}
\begin{eqnarray}
&&{\cal R}_{_{\mu\nu}}+\frac{1}{4}H_{_{\mu\rho\sigma}} {H^{^{\sigma \rho}}}_{_{\nu}}+2{\overrightarrow{\nabla}_{_\mu}}
{\overrightarrow{\nabla}_{_\nu}} \Phi ~=~0,\nonumber\\
&&(-1)^{^{\lambda}} {\nabla}^{^\lambda}\big(e^{^{-2 \Phi}} H_{_{\lambda\mu\nu}}\big)
~=~0,\nonumber\\
&&4 \Lambda-{\cal R}-\frac{1}{12} H_{_{\mu\nu\rho}}H^{^{\rho\nu\mu}} +4{\overrightarrow{\nabla}_{_\mu}} {\Phi} \overrightarrow{\nabla}^{^\mu} {\Phi}
 - 4 {\overrightarrow{\nabla}_{_\mu}} \overrightarrow{\nabla}^{^\mu} {\Phi}  =0,\label{2.24}
\end{eqnarray}
where the covariant derivatives ${\overrightarrow {\nabla}}_{_\mu}$, scalar curvature ${\cal R}$ and Ricci tensor ${\cal R}_{_{\mu\nu}}$
are calculated from the metric $G_{_{\mu\nu}}$ that is also used for lowering and raising
indices; moreover, ${\Phi}$  is the dilaton field which can be understood as an additional function on supermanifold $\M$
that defines the quantum non-linear $\sigma$-model and couples to scalar curvature of
the worldsheet.

The metric \eqref{2.22} is flat in the sense that its scalar curvature vanish.
One quickly finds that the only non-zero component of Ricci tensor
is ${\cal R}_{_{yy}}=-{1}/{2}$; and as the only non-zero components of $B$-filed are $B_{_{y \psi}}=\psi/2$ and
$B_{_{y \chi}}=\chi/2$, the only
non-zero components of the field strength $H$, which are obtained from \eqref{2.5}, are $H_{_{y \psi \psi}}=H_{_{y \chi \chi}} =1$.
Putting these pieces together, one verifies equations \eqref{2.24} with $\Lambda=0$ and
$\Phi =c_{_0} y +c_{_1}$ for some constants $c_{_0}$ and $c_{_1}$.

In order to investigate the super Poisson-Lie symmetry of the model on the one hand,
we need the left-invariant supervector fields on the $({C}_0^5 + {A})$ that by utilizing relation \eqref{2.20} we obtain
\begin{eqnarray}
V_{_1} &=& \frac{\overrightarrow{\partial}}{\partial y},\nonumber\\
V_{_2} &=& \frac{\overrightarrow{\partial}}{\partial x},\nonumber\\
V_{_3} &=&\frac{1}{2} \big(\psi \cos y-\chi \sin y\big) \frac{\overrightarrow{\partial}}{\partial x} + \cos y
~\frac{\overrightarrow{\partial}}{\partial \psi} -\sin y~ \frac{\overrightarrow{\partial}}{\partial \chi},\nonumber\\
V_{_4} &=& \frac{1}{2} \big(\psi \sin y+\chi \cos y\big) \frac{\overrightarrow{\partial}}{\partial x} + \sin y
~\frac{\overrightarrow{\partial}}{\partial \psi} +\cos y~ \frac{\overrightarrow{\partial}}{\partial \chi}. \label{2.25}
\end{eqnarray}
On the other hand, since the matrix background ${\cal E}_{_{\mu\nu}}$ is a composition of the metric ${G}_{_{\mu\nu}}$
and the $B$-field, it follows from relations \eqref{2.22} and \eqref{2.23} that
\begin{eqnarray}\label{2.26}
{\cal E}_{_{\mu\nu}}=\left( \begin{tabular}{cccc}
                $\beta$ & 1  &  $\psi$  & $\chi$\\
                1     & 0  &  0     & 0\\
                0    & 0  &  0  & 1\\
                0     & 0  & -1 & 0\\
                 \end{tabular} \right).
\end{eqnarray}
Substituting relations \eqref{2.25} and \eqref{2.26} into formula \eqref{2.11}
it is concluded that one cannot obtain a dual pair for the $({\C}_0^5 + {\A})$ Lie superalgebra that satisfies the condition
\eqref{2.11}. This means that the $({C}_0^5 + {A})$ WZW model does not include the super Poisson-Lie symmetry.

In summary, it has been shown that the WZW models based on the $GL(1|1)$ and $(C^3 + A)$ Lie supergroups have the super Poisson-Lie symmetry,
while this is not the case for the $(C_0^5 + A)$.
This however cannot be the end of the story since one may build
the gauged WZW models from Lie supergroups of the type $(2|2)$, even though these
do not include the super Poisson-Lie symmetry.
In the next section, in addition to building the gauged WZW models,
we address the question of whether there exists the super Poisson-Lie symmetry
for these gauged models so that we can find their corresponding dual pairs.
The physical motivation is clear. The discovery of these models represents
another step in the classification of exact conformal field theories of superdimension $(1|2)$ of WZW type.


\section{The $G/H$ gauged WZW models from Lie supergroups of the type $(2|2)$}
\label{Sec.3}

There are certain restrictions as to what subgroups of the isometry supergroup $G \times G$ we can gauge.
In order to examine this issue and following \cite{OFarrill},
let us assume that $G$ is a Lie supergroup and $\G$ its Lie superalgebra,
and $H \subset G \times G$ be a Lie subgroup with Lie algebra ${\mathfrak h}$ which
is spanned by the bases $\{t_i\}$.
We define $\big<. ~, ~.\big>_{_G}$ and $\big<. ~, ~.\big>_{_H}$ as the ad-invariant inner products on the respective Lie superalgebras.
The embedding ${\mathfrak h} \subset  \G \times \G$ defines two Lie algebra homomorphisms $l, r: {\mathfrak h} \rightarrow \G $ by composing with the cartesian projections. Then $H$ can be gauged if and only if the following condition holds
\begin{eqnarray}\label{3.0}
\big<l(t_i) , l(t_j)\big>_{_G} = \big<r(t_i) , r(t_j)\big>_{_G} =\big<t_i , t_j\big>_{_H},~~~~~~~\forall t_i, t_j \in {\mathfrak h},
\end{eqnarray}
such subgroups are called anomaly-free.
Provided that the consistency requirement \eqref{3.0}
is satisfied, these data define a conformally invariant $\sigma$-model on $G/H$ supercoset space via the construction of gauged WZW models
\cite{OFarrill} (see, also, \cite{cosetHeisenberg}).

Before we proceed to build the gauged WZW models from Lie supergroups of the type $(2|2)$,
let us turn our attention to the model setting.
We concentrate on the case in which $G$ is a Lie supergroup.
We are interested in gauging a one-dimensional subgroup $H$ of the symmetry group of WZW model action \eqref{2.2},
with the gauge transformation $g \rightarrow  e^{\varepsilon^i l(t_i)}  g  e^{\varepsilon^i r(t_i)}$, where $\varepsilon^i$ are the worldsheet dependent parameters. One may make this global symmetry local by
introducing gauge fields $A_{\pm}$ which take values in bosonic subgroup $H$ such that
$A_\pm = {{\mathbb{A}}}^i_\pm ~t_i$. If ${\epsilon} =\varepsilon^i~ t_i$ is an
infinitesimal gauge parameter, then the local axial symmetry is generated by
\begin{eqnarray}\label{3.1}
\delta g = { \epsilon} g + g { \epsilon},~~~~~~~~\delta {{\mathbb{A}}}^i_\pm = - \partial_{\pm} \varepsilon^i.
\end{eqnarray}
This local axial symmetry is a symmetry of the following gauged WZW action
\begin{eqnarray}\label{3.2}			
S[g, A_{_\pm}] &=& S_{_{WZW}}(g)+\int_{_\Sigma}d\sigma^+ d\sigma^-  \Big[\big<A_+~ ,~g^{-1} \partial_- g\big> + \big<A_-~ ,~ \partial_+ g g^{-1}\big>~~~~~~
\nonumber\\
&&~~~~~~~~~~~~~~~~~~~~~~~~~~~~~~~~~~~~+\big<A_-~ ,~A_+\big> + \big<A_-~ ,~g A_+ g^{-1}\big>\Big],
\end{eqnarray}
where the first term on the right side of \eqref{3.2} is given by \eqref{2.2}.
In what follows, we shall obtain three new conformal field theories or supercoset models $G/H$
by gauging an anomaly-free subgroup H=SO(2) of the $GL(1|1)$, $(C^3 + A)$ and $({C}_0^5 + {A})$ Lie supergroups.

\subsection{The GL(1|1)/SO(2) gauged WZW model}

We now describe the conformal field theory construction of WZW type which yields new supergeometry of the type $(1|2)$.
The model is constructed on the supercoset GL(1|1)/SO(2).
The Lie superbrackets of $gl(1|1)$ has been given by equation \eqref{2.6}. First of all, we calculate the elements within action
\eqref{3.2}. Applying a parametrization of $GL(1|1)$ as in \eqref{2.8} we find that
\begin{eqnarray}
g^{-1} \partial_{_-} g &=& \partial_{_-} y~ T_{_1} +  (\partial_{_-} x -  \partial_{_-} \chi ~\psi e^y) T_{_2}+
(\partial_{_-} \psi +  \partial_{_-}y ~\psi)T_{_3} +\partial_{_-} \chi~e^y T_{_4},\label{3.3}\\
 \partial_{_+} g  g^{-1} &=& \partial_{_+} y~ T_{_1} +  (\partial_{_+} x +  \partial_{_+} \psi ~\chi e^y) T_{_2}+
 \partial_{_+} \psi e^y T_{_3} +(\partial_{_+} \chi +  \partial_{_+}  y~ \chi) T_{_4}. \label{3.4}
\end{eqnarray}
We then gauge the one-dimensional subgroup H=SO(2) generated by the base $\{t_i\}=T_1$ of
the $gl(1|1)$\footnote{Here we have chosen one particular axial gauging.
Now the question may arise why the subgroup defined by the base $T_{_1}$ was chosen?
The answer to this question is that for the WZW model on the $GL(1|1)$, where the one-dimensional subgroup $H$ is
generated by a non-null element $T_{_1} \in \G$, any embedding ${\mathfrak h} \in T_{_1}
\rightarrow (\lambda_{_1} T_{_1} ,  \lambda_{_2} T_{_1}) \in \G \oplus \G$, for
$\lambda_{_1}, \lambda_{_2} \in \mathbb{R}$, satisfies the condition in \eqref{3.0}.
Indeed, $\big<l(T_{_1}) , l(T_{_1})\big>_{_G} = \lambda_{_1}^2 \big<T_{_1} , T_{_1}\big>_{_H}= \lambda_{_1}^2 \beta$
and, similarly, $\big<r(T_{_1}) , r(T_{_1})\big>_{_G} = \lambda_{_2}^2 \big<T_{_1} , T_{_1}\big>_{_H} =\lambda_{_2}^2 \beta$.
But, moreover, since $T_{_1}$ is a non-null element for any $\beta \neq 0$, we will take $\lambda_{_1} =  \lambda_{_2}=1$ in our treatment.}.
Accordingly, we have
${{A}_\pm} = {\mathbb{A}_\pm}~ T_{_1}$ and ${\epsilon} = \varepsilon~ T_{_1}$. Using these, one quickly obtains
\begin{eqnarray}
g {{A}_{_+}} g^{-1}  = {\mathbb{A}}_{_+} (T_{_1} - \psi \chi e^y T_{_2}- \psi e^y T_{_3} + \chi T_{_4}).\label{3.5}
\end{eqnarray}
Note that here the inner product is defined by $\Omega_{ab}$ as in \eqref{2.7}. Thus, the full action is now
\begin{eqnarray}
S[g , A_\pm]&=&S_{_{WZW}}(g)+\int_{_\Sigma}d\sigma^+ d\sigma^- \Big[{\mathbb{A}_{_+}}  (\beta \partial_{_-} y + \partial_{_-} x -
\partial_{_-} \chi e^y \psi)\nonumber\\
&&~~~~~~~~~~~~~~+ {\mathbb{A}_{_-}}  (\beta \partial_{_+} y + \partial_{_+} x + \partial_{_+} \psi e^y \chi) +
{\mathbb{A}_+}  {\mathbb{A}_-}(2 \beta - \psi \chi e^y)\Big].\label{3.6}
\end{eqnarray}
Integrating over the gauge fields leads to
\begin{eqnarray}
{\mathbb{A}_{_+}}&=&-\frac{1}{2 \beta} (\beta \partial_+ y + \partial_+ x + \partial_+ \psi e^{ y}  \chi) (1+ \frac{1}{2 \beta} e^{ y} \psi\chi),\nonumber\\
{\mathbb{A}_{_-}}&=&-\frac{1}{2 \beta} (\beta \partial_- y + \partial_- x - \partial_- \chi e^{ y}  \psi) (1+ \frac{1}{2 \beta} e^{ y} \psi\chi).\label{3.7}
\end{eqnarray}
Inserting \eqref{3.7} into \eqref{3.6} and then utilizing \eqref{2.10} we arrive at
\begin{eqnarray}
S[g , A_{_\pm}] &=& S_{_{WZW}}(g)+\int_{_\Sigma}d\sigma^+ d\sigma^-  ~{\mathbb{A}_+}  {\mathbb{A}_-}(\psi \chi e^y -2 \beta)\nonumber\\
&=&\frac{1}{2}\int_{_\Sigma}d\sigma^+ d\sigma^-\Big[\beta \partial_+ y \partial_- y +
\partial_+ y \partial_- x+ \partial_+ x \partial_- y\nonumber\\
&&~~~~~~~~~~~~~~~~~~-2 e^{ y} \partial_+ \psi \partial_- \chi +  2 {\mathbb{A}_+}  {\mathbb{A}_-}(\psi \chi e^y -2 \beta)\Big].\label{3.8}
\end{eqnarray}
On the other hand, the action \eqref{3.8} must be invariant under the axial gauging transformations. For this purpose, by using \eqref{3.1}
one finds that
\begin{eqnarray}
\delta y =2 \varepsilon ,~~~\delta x =0,~~~~\delta \psi = -\varepsilon \psi,~~~~\delta \chi = -\varepsilon \chi,
 ~~~~\delta {\mathbb{A}_{_\pm}}= -{\partial_{_\pm}}\varepsilon.\label{3.9}
\end{eqnarray}
Indeed, the action \eqref{3.8} is invariant under the above transformations.
We can now gauge fix by setting $x=-\beta y$. After making this gauge choice and eliminating ${\mathbb{A}_{_\pm}}$, the action
becomes
\begin{eqnarray}
S[g , A_{_\pm}]=\frac{1}{2}\int_{_\Sigma}d\sigma^+ d\sigma^-\Big[- \beta \partial_+ y \partial_- y
-2 e^y  \partial_+ \psi \partial_- \chi + \frac{1}{\beta} \partial_+ \psi ~ \psi \chi e^y   \partial_- \chi\Big].\label{3.10}
\end{eqnarray}
The supersymmetric part of the action gives the metric,
whereas the super anti-symmetric part gives the tensor $B_{\mu \nu }$. Thus,
the corresponding line element and $B$-field in the coordinate basis $(y; \psi, \chi)$ are, respectively, read off
\begin{eqnarray}
ds^2 &=& -\beta dy^2 -2 (e^y- \frac{1}{2\beta} \psi \chi e^{2y}) d \psi d \chi,\nonumber\\
B &=& -(e^y- \frac{1}{2\beta} \psi \chi e^{2y})~ d\psi \wedge d \chi.\label{3.11}
\end{eqnarray}
As a gauged WZW model, this model should be conformally invariant\footnote{As before, gauged WZW models are conformally invariant to all oders by means of the GKO construction (under very minor assumptions) \cite{GKO}. In fact, there is no need to check this explicitly on a one-loop level.
Just to find the corresponding dilaton field we look at equations \eqref{2.24}.}.
In order to find the dilaton field corresponding to background \eqref{3.11},
we first look at the one-loop beta function equations, \eqref{2.24}.
One quickly finds that the scalar curvature of the metric is ${\cal R}= (\psi \chi e^y -5 { \beta})/{2 \beta^2}$, and
the only non-zero components of the field strength are $H_{_{y \psi \chi}}=(\beta e^y- \psi \chi e^{2y})/ \beta$,
$H_{_{\psi \psi \chi}}= e^{2y} \chi/ \beta$ and $H_{_{\psi \chi \chi}}=-e^{2y} \psi/ \beta$.
Putting these pieces together, one verifies equations \eqref{2.24} with $\Lambda =0$ and dilaton field
$\Phi= c_{_0}+ \frac{1}{4 \beta} \psi \chi e^y$ for some constant $c_{_0}$.

In order to add some interpretation of the gauged background, it is useful to look at the isometry symmetries of the metric.
To this end, one must calculate the Killing supervectors corresponding to the metric of supercoset geometry.
In this way, we use the graded form of Killing equation,
\begin{eqnarray}\label{1}
{\cal L}_{_{K_a}} {G}_{_{\mu \nu}} = (-1)^{\mu +\lambda +\mu a} \frac{\overrightarrow{\partial} {K_a}^{\lambda}}{{\partial} x^{^{\mu}}} {G}_{\lambda \nu}+ {K_a}^{\lambda} \frac{\overrightarrow{\partial} {G}_{\mu\nu}}{{\partial} x^{^{\lambda}}}
+ (-1)^{\mu \nu +\mu\lambda + \lambda+ \nu a +\nu} \frac{\overrightarrow{\partial} {K_a}^{\lambda}}{{\partial} x^{^{\nu}}} {G}_{\mu\lambda} = 0.
\end{eqnarray}
where $K_a={K_a}^{\mu} \frac{\overrightarrow{\partial}}{{\partial} x^{^{\mu}}}$ stands for the Killing supervector.
Using \eqref{1}, it can be shown that the $GL(1|1)/SO(2)$ metric admits only the following four bosonic Killing vectors
\begin{align}\label{1.1}
K^B_{_1} = \psi \overrightarrow{\frac{\partial}{\partial \psi}}-\overrightarrow{\frac{\partial}{\partial y}},~~~~~~
K^B_{_2} = - \overrightarrow{\frac{\partial}{\partial y}} +\chi \overrightarrow{\frac{\partial}{\partial \chi}},~~~~~~
K^B_{_3} =  \chi \overrightarrow{\frac{\partial}{\partial \psi}},~~~~~~~
K^B_{_4} = \psi \overrightarrow{\frac{\partial}{\partial \chi}}.
\end{align}
We note that there is no fermionic Killing vector for the desired metric.

\subsection{The $(C^3 + A)/SO(2)$ gauged WZW model}
\label{subsec.3.2}

In order to write down the action \eqref{3.2} on the supercoset $(C^3 + A)/$SO(2)
we need to calculate the left- and right-invariant super one-forms on the $(C^3 + A)$.
The corresponding left-invariant super one-form has been obtained by equations \eqref{2.8} and \eqref{2.12} in section \ref{Sec.2}. In the same way,
one gets
\begin{eqnarray}
\partial_{_+} g  g^{-1} = \partial_{_+} y~ T_{_1} +  (\partial_{_+} x +  \partial_{_+} \chi ~\frac{\chi}{2}) T_{_2}+
(\partial_{_+} \psi -\partial_{_+} y~\chi) T_{_3} +  \partial_{_+}  \chi T_{_4}. \label{3.12}
\end{eqnarray}
Analogously, we gauge the one-dimensional subgroup H=SO(2) generated by $T_{_1}$ of the $({\C}^3 + \A)$.
Accordingly, we have  ${{A}_\pm} = {\mathbb{A}_\pm} T_1$ and ${\epsilon} = \varepsilon  T_1$. Using these together with \eqref{3.1} one can find
the axial gauging transformations, giving us
\begin{eqnarray}
\delta y =2 \varepsilon,~~~\delta x =0,~~~~\delta \psi = \varepsilon \chi,~~~~\delta \chi = 0,
 ~~~~\delta {\mathbb{A}_\pm}= -{\partial_\pm} \varepsilon.\label{3.13}
\end{eqnarray}
Considering the ungauged WZW action from equation \eqref{2.14}, the gauged WZW action in this case is
\begin{eqnarray}
S[g , A_{_{\pm}}]&=&\frac{1}{2}\int_{_\Sigma}d\sigma^+ d\sigma^- \Big[\beta\partial_{_{+}} y \partial_{_{-}} y+\partial_{_{+}} y \partial_{_{-}} x+ \partial_{_{+}} x \partial_{_{-}} y-\partial_{_{+}} \psi \partial_{_{-}} \chi+\partial_{_{+}} \chi \partial_{_{-}} \psi
+\partial_{_{+}} y  \chi \partial_{_{-}} \chi\nonumber\\
&+&2 {\mathbb{A}_{_+}} (\beta\partial_{_{-}} y + \partial_{_{-}} x-\partial_{_-} \chi ~\frac{\chi}{2})+
2 {\mathbb{A}_{_-}} (\beta\partial_{_{+}} y + \partial_{_{+}} x+\partial_{_+} \chi ~\frac{\chi}{2})+4 \beta {\mathbb{A}_{_+}}  {\mathbb{A}_{_-}}\Big].~~~~~~\label{3.14}
\end{eqnarray}
This action is invariant under the transformations given in \eqref{3.13}.
One can obtain the $\sigma$-model by fixing the gauge and then integrate over the gauge
fields. A convenient gauge choice is $x+\beta y =0$ and the resulting $\sigma$-model action is
\begin{eqnarray}
S[g , A_{_{\pm}}]=\frac{1}{2}\int_{_\Sigma}d\sigma^+ d\sigma^-\Big[- \beta \partial_+ y \partial_- y
+ \partial_+ y  \chi \partial_- \chi - \partial_+ \psi ~ \partial_- \chi + \partial_+ \chi ~ \partial_- \psi\Big].\label{3.15}
\end{eqnarray}
The line element and $B$-field corresponding to the above action are
\begin{eqnarray}
ds^2&=&-\beta dy^2 + \chi dy d\chi -2 d \psi d\chi,\nonumber\\
B&=&\frac{\chi}{2}~ dy \wedge d \chi.\label{3.16}
\end{eqnarray}
The metric is flat in the sense that its Ricci tensor and scalar curvature vanish;
moreover, only the non-zero component of the field strength corresponding to the $B$-field is easily obtained to be $H_{y \chi \chi} =1$. Thus,
one quickly concludes that background \eqref{3.16} with a constant dilaton field and zero cosmological
constant satisfy the vanishing of the one-loop beta function equations.

Comparing the background \eqref{3.16} with the previous one, it can be noted that
the $(C^3 + A)/SO(2)$ metric has more symmetries than the $GL(1|1)/SO(2)$ one.
Using the formula \eqref{1}, one can show that the $(C^3 + A)/SO(2)$ metric admits four bosonic Killing vectors
\begin{align}\label{2}
K^B_{_1} &= \overrightarrow{\frac{\partial}{\partial y}},~~~~~~~~~~~~~~~~~~~~~~~~~~~~~~~~
K^B_{_2} = (\psi-\frac{y \chi}{2})\Big[\frac{y}{2} \overrightarrow{\frac{\partial}{\partial \psi}} + \overrightarrow{\frac{\partial}{\partial \chi}}\Big],\nonumber\\
K^B_{_3} &= (\psi-{y \chi}) \overrightarrow{\frac{\partial}{\partial \psi}} -\chi \overrightarrow{\frac{\partial}{\partial \chi}},~~~~~~~~~~
K^B_{_4} = \chi \overrightarrow{\frac{\partial}{\partial \psi}}.
\end{align}
One can easily check that the Lie algebra spanned by these four bosonic vectors is the $gl(2, \mathbb{R})$.
In addition, there exist four fermionic Killing vectors $(K^F_{_1}, K^F_{_2}, K^F_{_3}, K^F_{_4})$ which
generate the isometry Lie superalgebra of the metric together with the $gl(2, \mathbb{R})$,
\begin{align}\label{3}
K^F_{_1} &=(\psi-\frac{y \chi}{2}) \overrightarrow{\frac{\partial}{\partial y}}+ \frac{1}{2} (\beta y^2 + \psi \chi) \overrightarrow{\frac{\partial}{\partial \psi}} + \beta y \overrightarrow{\frac{\partial}{\partial \chi}},~~~~~~~~~K^F_{_2} = \overrightarrow{\frac{\partial}{\partial \psi}},\nonumber\\
K^F_{_3} &= {\chi} \overrightarrow{\frac{\partial}{\partial y}} -\beta y  \overrightarrow{\frac{\partial}{\partial \psi}},
~~~~~~~~~~~~~~~~~~~~~~~~~~~~~~~~~~~~~~~~~~~~
K^F_{_4} = \frac{y}{2} \overrightarrow{\frac{\partial}{\partial \psi}} + \overrightarrow{\frac{\partial}{\partial \chi}}.
\end{align}

\subsection{The $(C_0^5 + A)/SO(2)$ gauged WZW model}

To construct the $(C_0^5 + A)/$SO(2) gauged WZW model, we apply the parametrization \eqref{2.18} of the
$(C_0^5 + A)$ supergroup. We need to calculate $\partial_{_+} g  g^{-1}$, which gives us
\begin{eqnarray}
\partial_{_+} g  g^{-1} &=& \partial_{_+} y~ T_{_1} +  (\partial_{_+} x + \partial_{_+} y~ \psi \chi+\partial_{_+} \psi ~\frac{\psi}{2} + \partial_{_+} \chi ~\frac{\chi}{2}) T_{_2}\nonumber\\
&& ~~~~~~~~~~~~~+(\partial_{_+} \psi -\partial_{_+} y~\chi) T_{_3} +  (\partial_{_+} \chi +\partial_{_+} y~\psi) T_{_4}. \label{3.17}
\end{eqnarray}
Similarly to previous cases, we gauge the one-dimensional subgroup H=SO(2) generated by $T_1$ of the $({\C}_0^5 + \A)$.
Applying \eqref{2.21} and using  \eqref{3.2}, the gauged WZW action is written as
\begin{eqnarray}
S[g , A_{_{\pm}}]
&=&\frac{1}{2}\int_{_\Sigma}d\sigma^+ d\sigma^- \Big[\beta\partial_{_{+}} y \partial_{_{-}} y+\partial_{_{+}} y \partial_{_{-}} x+ \partial_{_{+}} x \partial_{_{-}} y
+\partial_{_{+}} y  \psi  \partial_{_{-}} \psi+\partial_{_{+}} y  \chi  \partial_{_{-}} \chi\nonumber\\
~~~~~~~~&& -\partial_{_{+}} \psi ~ \partial_{_{-}} \chi +\partial_{_{+}} \chi ~ \partial_{_{-}} \psi +2 {\mathbb{A}_{_+}}(\beta \partial_{_{-}} y+\partial_{_{-}} x -\partial_{_-} \psi ~\frac{\psi}{2} -\partial_{_-} \chi ~\frac{\chi}{2})\nonumber\\
 &&+2 {\mathbb{A}_{_-}}(\beta \partial_{_{+}} y+\partial_{_{+}} x +\partial_{_+} \psi ~\frac{\psi}{2} +\partial_{_+} \chi ~\frac{\chi}{2}
+\partial_{_{+}} y  ~\psi \chi) +2 {\mathbb{A}_{_+}}  {\mathbb{A}_{_-}} (2 \beta + \psi\chi)\Big],~~~~~~~~~\label{3.18}
\end{eqnarray}
and it is invariant under the transformations
\begin{eqnarray}
\delta y =2 \varepsilon,~~~\delta x =0,~~~~\delta \psi = \varepsilon \chi,~~~~\delta \chi =- \varepsilon \psi,
 ~~~~\delta {\mathbb{A}_\pm}= -{\partial_\pm} \varepsilon.\label{3.19.2}
\end{eqnarray}
We can now gauge fix by setting $x+\beta y=0$.
After making this gauge choice and eliminating ${\mathbb{A}_{_\pm}}$ the action becomes
\begin{eqnarray}
S[g , A_{_{\pm}}]&=&\frac{1}{2}\int_{_\Sigma}d\sigma^+ d\sigma^-\Big[- \beta \partial_+ y \partial_- y
+ \partial_+ y  \psi \partial_- \psi + \partial_+ y  \chi \partial_- \chi\nonumber\\
~~~~~&&- \partial_+ \psi ~ \partial_- \chi+ \partial_+ \chi ~ \partial_- \psi  + \frac{1}{4 \beta}
(-\partial_+ \psi ~\psi\chi  \partial_- \chi + \partial_+ \chi \psi\chi ~\partial_- \psi)\Big].\label{3.19}
\end{eqnarray}
By identifying the gauged WZW action above with the $\sigma$-model action \eqref{2.4}
one can read off the metric and the super anti-symmetric tensor.
The corresponding line element and the $B$-field are therefore
\begin{eqnarray}
ds^2&=&-\beta dy^2 + \psi dy d \psi +\chi dy d\chi -2 (1+ \frac{\psi \chi }{4 \beta}) d \psi d\chi,\nonumber\\
B&=&\frac{1}{2}~(\psi dy \wedge d\psi + \chi dy \wedge d \chi).\label{3.20}
\end{eqnarray}
One can verify that the scalar curvature of the metric is ${\cal R} =(5 \beta+ \psi \chi)/ 2 \beta^2$,
and only the non-zero components of the field strength
are $H_{y \psi \psi} = H_{y \chi \chi} =1$ and thus, the vanishing of beta function equations are
indeed satisfied with $\Lambda =0$ and dilaton field $\Phi= c_{_0}- \frac{1 }{4 \beta} \psi \chi$
for some constant $c_{_0}$.
The properties of metric of $(C^5_0 + A)/SO(2)$ are very similar to that of the $GL(1|1)/SO(2)$ geometry,
as both geometries have scalar curvature and dilaton field dependent on $\psi \chi$.
It is also interesting to note that the $(C^5_0 + A)/SO(2)$ metric also admits only four bosonic Killing vectors, just like the $GL(1|1)/SO(2)$ metric.

In summary, in this section we gauged the subgroup SO(2) generated by the $T_{_1}$ inside each of the $(2|2)$-dimensional Lie supergroups, and
obtained three $(1|2)$-dimensional gauged WZW models.
Since the bosonic part of the resulting models was one-dimensional, finding a physical interpretation for them was not an easy task.
In fact, a $(1|2)$-dimensional model is too small for realistic phenomenology, and a four-dimensional spacetime is needed to study this issue.
The reason for the choice of the subgroup SO(2) was also explained in footnote 12.
Before closing this section, some questions may arise: By taking into account formula \eqref{3.0},
are there other Abelian subgroups that are anomaly-free? Could one gauge a different linear combination or consider the vector gauging instead
of axial? Before answering these questions we note that the vector gauged WZW model is defined in the same way as the axial model of \eqref{3.2},
except that only the sign behind terms $\big<A_+~ ,~g^{-1} \partial_- g\big>$ and $\big<A_-~ ,~g A_+ g^{-1}\big>$ is negative.
If one begins to build a vector gauging of the $(C^3 + A)/SO(2)$ WZW model by gauging the subgroup generated by
the $T_{_1}$, then he/she concludes that $\big<A_-~ ,~A_+\big> - \big<A_-~ ,~g A_+ g^{-1}\big> =0$.
Thus, the term ${\mathbb{A}_+} {\mathbb{A}_-}$ will not appear in the gauged action
so that we can determine both the ${\mathbb{A}_+}$ and $ {\mathbb{A}_-}$ by integrating over the gauge fields.
Instead, this problem will not exist for building a vector gauging of the $GL(1|1)/SO(2)$ and $(C_0^5 + A)/SO(2)$ WZW models.
Because for them one obtains that
$\big<A_-~ ,~A_+\big> - \big<A_-~ ,~g A_+ g^{-1}\big> = {\mathbb{A}_+} {\mathbb{A}_-} \psi \chi e^y $ and $\big<A_-~ ,~A_+\big> - \big<A_-~ ,~g A_+ g^{-1}\big> = - {\mathbb{A}_+} {\mathbb{A}_-} \psi \chi$, respectively.
However, we want to clarify that other gauging is possible or relevant besides $T_{_1}$-gauging.

If one performs the gauging by a null element $T_{_2}$ of the $SO(2)$ subgroup inside each of the $(2|2)$ Lie supergroups,
then we have that ${{A}_\pm} = {\mathbb{A}_\pm} T_2$ and ${\epsilon} = \varepsilon  T_2$.
Furthermore, this choice of gauging satisfies condition \eqref{3.0}. Since the $T_{_2}$ is null, $\big<T_{_2}~ ,~T_{_2}\big> =0$,
we find that $\big<A_-~ ,~A_+\big> = \big<A_-~ ,~g A_+ g^{-1}\big> =0$. Therefore, we cannot build the vector and axial gauged WZW models
by the $T_{_2}$ inside each of the $(2|2)$ Lie supergroups.
On the other hand, as shown the WZW models on each of the $(2|2)$ Lie supergroups involve two bosonic fields and two fermionic ones.
Accordingly, gauging with a linear combination of the two bosonic bases, ${{A}_\pm} = {\mathbb{A}_\pm^1} T_1+{\mathbb{A}_\pm^2} T_2$, is not possible,
because even if one could perform the gauging process with a two-dimensional Abelian subgroup $SO(2) \times SO(2)$ generated
by $(T_{_1}, T_{_2})$, then we find that the metric of the gauged model only includes two fermionic fields, which will no longer be superinvertible.

\section{An explicit example of super Poisson-Lie symmetric gauged WZW models:  The case of $(C^3 + A)/SO(2)$}
\label{Sec.4}

In this section we shall show that the supercoset $\sigma$-model on the $(C^3 + A)/$SO(2) (derived in subsection \ref{subsec.3.2}) has the
super Poisson-Lie symmetry. In this way, we construct a dual pair for the background on the supercoset $(C^3 + A)/$SO(2) by
applying the super Poisson-Lie T-duality on the Drinfeld superdouble $\big((A_{1,1} +2A)^0 , C^3\big)$ \cite{ER4}.
By using a certain parametrization of the $(A_{1,1} +2A)^0$ Lie supergroup and
by a suitable choice of constant matrix $E_0(e)$ we construct the original $\sigma$-model including background on the $(C^3 + A)/$SO(2).
Before proceeding to do this, let us review the construction of
super Poisson-Lie T-dual $\sigma$-models on Lie supergroups \cite{{ER2},{ER5}}.


\subsection{A review of super Poisson-Lie T-duality without spectator fields}

Both the original and dual geometries of the super Poisson-Lie T-dualizable $\sigma$-models are derived from
the so-called Drinfeld superdouble. The Drinfeld superdouble of a Lie supergroup $G$ is defined as a Lie supergroup $D$, with superdimension twice the
one of $G$, such that its Lie superalgebra $\D$ can be decomposed into a pair of maximally isotropic sub-superalgebras,
$\G$ and $\tilde \G$ with respect to a non-degenerate invariant bilinear form on $\D$, with  $\G$ and $\tilde \G$ respectively
the Lie superalgebra of $\G$ and its dual superalgebra. The dual superalgebra is endowed with a Lie superbracket which
has to be compatible with existing structures.
The construction of Poisson-Lie T-dual $\sigma$-models on Lie groups has been described
in \cite{{Klim1},{Klim2}}. Then, this construction was generalized to the super case in \cite{{ER2},{ER5}}.
The models have target supermanifolds as the Lie supergroups $G$ and ${\tilde G}$ and are, respectively,  given by the actions
\begin{eqnarray}
S&=&\frac{1}{2}\int_{_{\Sigma}}\!d\sigma^{+}
d\sigma^{-}\;(-1)^{^{a}} {{R_{_+}}}^{\hspace{-1.5mm} a}\;  {E_{_{ab}}}(g)\;{{R_{_-}}}^{\hspace{-1.5mm} b},\label{4.1}\\
{\tilde  S}&=&\frac{1}{2}\int_{_{\Sigma}}\!d\sigma^{+}
d\sigma^{-}\;(-1)^{^{b}} {{\tilde  R_{_+ a}}}\;  {{\tilde E}^{^{ab}}}(\tilde g)\;{{\tilde  R_{_- b}}},\label{4.2}
\end{eqnarray}
where ${{R_{_\pm}}}^{\hspace{-1.5mm} a}$ and ${{\tilde  R_{_\pm a}}}$ are the components of the right-invariant super one-forms on the
$G$ and $\tilde G$, respectively,
which are defined by means of the elements $g: \Sigma \rightarrow G$ and $\tilde g: \Sigma \rightarrow \tilde G$ in the following forms
\begin{eqnarray}
\partial_{_{\pm}} g  ~g^{-1}&=&(-1)^a ~{R_{_{\pm}}^{a}}~T_{{_a}}=(-1)^a ~\partial_{_\pm} x^{^{\mu}}
\; {{_{_\mu}} R}^{^a}~T_{{_a}}, \label{4.3}\\
\partial_{\pm}{\tilde g} ~{\tilde g}^{-1} &=&{{\tilde  R_{_\pm a}}}~ {\tilde  T}^{{^a}}= \partial_{\pm} {\tilde x}^\mu~ {{_{_\mu}} \tilde R}_{a} {\tilde  T}^{{^a}}.\label{4.4}
\end{eqnarray}
The background fields $E_{_{ab}}(g)$ and ${{\tilde E}^{^{ab}}}(\tilde g)$ are defined
by\footnote{In order to calculate the superinverse of the matrices one must use the superinverse formula introduced in \cite{D}.}
\begin{eqnarray}\label{4.5}
E(g) = \left(E^{-1}_0(e) + \Pi (g)\right)^{-1},~~~~{\tilde E}(\tilde g) = (E_0(e) + {\tilde \Pi} (\tilde g))^{-1},
\end{eqnarray}
where $E_0(e)$ is the $\sigma$-model constant matrix at the unit element of $G$.
The $\Pi(g)$ defined by $\Pi^{ab}(g)=(-1)^c~ b^{^{ac}}(g)  {(a^{-1})_c}^{b} (g)$ is the super Poisson structure on the $G$, in which
$a(g)$  and $b(g)$ are sub-matrices of the adjoint representation of the supergroup $G$ on
$\D$ in the basis $(T_{{_a}} , {\tilde T}^{{^a}})$, which are defined
\begin{eqnarray}\label{4.6}
g^{-1} T_{{_a}}~ g &=&(-1)^c ~a_{_{a}}^{^{~c}}(g) ~ T_{{_c}},\nonumber\\
g^{-1} {\tilde T}^{{^a}} g &=&
(-1)^c ~b^{^{ac}}(g)~ T_{{_c}}+{(a^{^{-st}})^{{~a}}}_{c}(g)~{\tilde T}^{{^c}}.
\end{eqnarray}
Notice that the super Poisson structure on the $\tilde G$, ${\tilde \Pi}_{ab}(\tilde g)$,
is defined as in $\Pi^{ab}(g)$ by replacing untilded quantities with tilded ones.

\subsection{The original $\sigma$-model: background on the $(C^3 + A)/SO(2)$}

As mentioned above, we shall obtain the background on $(C^3 + A)/$SO(2) from a T-dualizable $\sigma$-model
constructing on the superdouble $\big((A_{1,1} + 2A)^0 , C^3\big)$.
The $({\A}_{1,1} +2{\A})^0$ and ${\C}^3$ are three-dimensional Lie superalgebras of the type $(1 | 2)$ \cite{{B},{ER4}}.
They are spanned by the set of generators $\{T_{_1}; T_{_2}, T_{_3}\}$  and $\{{\tilde T}^{^1}; {\tilde T}^{^2}, {\tilde T}^{^3}\}$, respectively.
The six-dimensional Lie superalgebra of the superdouble $\big(({\A}_{1,1} +2{\A})^0 , {\C}^3\big)$ is defined by
the following non-zero (anti)commutation relations\footnote{ $(T_{_1},  {\tilde T}^{^1})$ and
$(T_{_2}, T_{_3}, {\tilde T}^{^2}, {\tilde T}^{^3})$ are bosonic and fermionic bases, respectively.}:
\begin{eqnarray}\label{4.7}
\{T_3 , T_3\}=\frac{1}{\beta}T_1,~~[{\tilde T}^1 , {\tilde T}^2]= \frac{1}{2 \beta} {\tilde T}^3,~~~
[T_3 , {\tilde T}^1]=\frac{1}{2\beta}T_2 -\frac{1}{\beta} {\tilde T}^3,~ ~ \{T_3 , {\tilde T}^2\}=\frac{1}{2\beta}T_1,
\end{eqnarray}
for a non-zero constant $\beta$.
In order to write the action of the original $\sigma$-model explicitly we need to find the components of the right-invariant super one-forms
on the $(A_{1,1} + 2A)^0$. To this purpose we use the following parametrization of the group supermanifold:
\begin{eqnarray}\label{4.8}
g~=~e^{\chi T_3} ~e^{y T_1}~e^{\psi T_2},
\end{eqnarray}
where $y$ is a bosonic field, while $(\psi, \chi)$ are fermionic ones. Using \eqref{4.3} and \eqref{4.7} together with \eqref{4.8} one gets
\begin{eqnarray}\label{4.9}
R_{\pm}^1= \partial_{\pm} y + \frac{1}{2\beta} \partial_{\pm} \chi ~ \chi,~~~~~~R_{\pm}^2 = -\partial_{\pm} \psi,~~~~~~
{R_{\pm}^3}=- \partial_{\pm} \chi.
\end{eqnarray}
For our purpose it is also necessary to compute the super Poisson structure.
Using equations \eqref{4.6}-\eqref{4.8} we get
\begin{eqnarray}\label{4.10}
\Pi^{^{ab}}(g) =\left( \begin{array}{ccc}
                     0 & \frac{\chi}{2\beta}  & 0\\

                     -\frac{\chi}{2\beta} & 0 & 0\\

                      0 & 0 & 0
                      \end{array} \right).
\end{eqnarray}
Let us now choose the $\sigma$-model constant matrix in the form of
\begin{eqnarray}\label{4.11}
{E^{-1}_0}^{ab}(e)=\left( \begin{array}{ccc}
                    -\frac{1}{\beta} & 0 & 0\\
                    0 & 0 & 1\\
                    0 & -1 & 0
                      \end{array} \right).
\end{eqnarray}
Inserting \eqref{4.10} and \eqref{4.11} into the first equation of \eqref{4.5} and then
utilizing $R_{\pm}^a$'s of equation \eqref{4.9} together with formula \eqref{4.1},
the original $\sigma$-model is worked out to be
\begin{eqnarray}\label{4.12}
S=\frac{1}{2}\int_{_\Sigma}d\sigma^+ d\sigma^-\Big[- \beta \partial_+ y \partial_- y
+ \partial_+ y  \chi \partial_- \chi - \partial_+ \psi ~ \partial_- \chi + \partial_+ \chi ~ \partial_- \psi\Big].
\end{eqnarray}
Indeed, this model is nothing but the gauged WZW model on the supercoset $(C^3 + A)/$SO(2) which was obtained in equation \eqref{3.15}
of subsection \ref{subsec.3.2}.

\subsection{The dual $\sigma$-model}
\label{subsec.4.3}
In order to find the dual pair for the model \eqref{4.12} we parameterize the $C^3$ Lie supergroup
with coordinates $({\tilde x}; {\tilde  \psi}, {\tilde  \chi})$ so that its element
is defined as in \eqref{4.8} by replacing untilded quantities with tilded ones.
 Accordingly, one may apply relation \eqref{4.4} to obtain the corresponding right-invariant super one-forms, giving us
\begin{eqnarray}\label{4.13}
{{\tilde  R_{_\pm 1}}}= \partial_{\pm} {\tilde  y },
~~~~~~{{\tilde  R_{_\pm 2}}} = \partial_{\pm} {\tilde  \psi},~~~~~~{{\tilde  R_{_\pm 3}}}=  \frac{1}{2\beta} \partial_{\pm} {\tilde  \psi} ~ {\tilde  y}
+\partial_{\pm} {\tilde  \chi}.
\end{eqnarray}
Utilizing relation \eqref{4.6} for tilded quantities we get
\begin{eqnarray}\label{4.14}
{\tilde \Pi}_{ab}(\tilde g) =\left( \begin{array}{ccc}
                      0 & 0 & 0\\

                      0 & 0 & 0\\

                      0 & 0 & -\frac{\tilde y}{\beta}
                      \end{array} \right).
\end{eqnarray}
Inserting \eqref{4.11} and \eqref{4.14} into the second relation of equation \eqref{4.5},
and using \eqref{4.13} and \eqref{4.2},  we can obtain the dual $\sigma$-model.
It is then read
\begin{eqnarray}\label{4.15}
{\tilde S}=\frac{1}{2}\int_{_\Sigma}d\sigma^+ d\sigma^-\Big[-\frac{1}{\beta} \partial_+ {\tilde y} \partial_- {\tilde y}
-\frac{1}{\beta} {\tilde y} \partial_+ {\tilde \psi} \partial_- {\tilde \psi}- \partial_+ {\tilde \psi}   \partial_- {\tilde \chi}
+ \partial_+ {\tilde \chi} ~ \partial_- {\tilde \psi}\Big].
\end{eqnarray}
Identifying the above action with the $\sigma$-model action
of the form \eqref{2.4} we can read off the background of model including
the line element, ${\tilde {ds}}^2$, and $\tilde B$-field in the coordinate base $(d \tilde x; d \tilde \psi, d  \tilde \chi)$ as
\begin{eqnarray}
{\tilde {ds}}^2 &=& -\frac{1}{\beta} d {\tilde y}^2 -2d {\tilde \psi} d {\tilde \chi},\label{4.16}\\
\tilde B &=& -\frac{1}{2\beta} \tilde y d {\tilde \psi} \wedge d {\tilde \psi}.\label{4.17}
\end{eqnarray}
The metric is flat in the sense that its scalar curvature and Ricci tensor vanish, ${\tilde {\cal R}}=0$, ${\tilde {\cal R}_{\mu \nu}}=0$.
In order to check the conformal invariance of the dual model we find that the only
non-zero component of the field strength corresponding to $\tilde B$-field \eqref{4.17}
is ${\tilde H}_{_{{\tilde y} {\tilde \psi} {\tilde \psi}}} =1 / \beta$. Then,
one verifies the one-loop beta function equations, \eqref{2.24}, with $\tilde \Lambda=0$ and constant dilaton field,
$\tilde \Phi =c_{_0}$.

To sum up, we have found an explicit example of super Poisson-Lie symmetric gauged WZW models which has the super Poisson-Lie symmetry. We were able to construct
the gauged WZW model on the supercoset $(C^3 + A)/$SO(2) from a T-dualizable $\sigma$-model
on the superdouble $\big((A_{1,1} +2A)^0 , C^3\big)$.
Moreover, we found a dual pair for the gauged WZW model so that it is conformally invariant at the one-loop order.
Indeed, this example is very worthwhile in its own right.


\section{A review of YB deformation of the WZW model on Lie supergroups}
\label{Sec.5}

As mentioned in the Introduction section, YB deformations of the $GL(1|1)$ and $(C^3 + A)$  WZW models
have been carried out by making use of the super skew-symmetric classical r-matrices satisfying (m)GCYBE \cite{Epr2}. Two of us in \cite{Epr2} showed that:

\begin{itemize}

\item any r-matrix of the $gl(1|1)$ Lie superalgebra as a solution of the (m)GCYBE belongs
just to the eleven inequivalent classes.
Then, using these inequivalent r-matrices, it was shown that the YB deformations of
the $GL(1|1)$ WZW model, including the metric and the $B$-field, are classified into the eleven families.

\item also, in the case of $({\C}^3 +{\A})$ Lie superalgebra, it was shown that
any r-matrix of the $({\C}^3 +{\A})$ as a solution of the (m)GCYBE belongs just to eight inequivalent classes.
In this way, it was obtained eight YB deformed model based on the $(C^3 + A)$.

\item in addition to these, by checking the conformal invariance of the models up to one-loop order,
it was concluded that the $GL(1|1)$ and $(C^3 + A)$ WZW models were conformal theories within the classes of the YB deformations preserving the conformal invariance.

\end{itemize}

Similarly, we shall construct the YB deformation of WZW model based on the $(C_0^5 + A)$.
Before proceeding to do this, let us review the setting related to the YB deformation of WZW model on Lie supergroups.
Inspired by a prescription invented by Delduc, Magro and Vicedo \cite{Delduc4}, it was
generalized \cite{Epr2} the YB deformation of WZW model from Lie groups to Lie supergroups.
The action of the YB deformed WZW model on a Lie supergroup $G$ may be expressed as
\begin{eqnarray}\label{5.1}			
S^{^{YB }}_{_{WZW}}(g)=\frac{1}{2}\int_{_\Sigma}d\sigma^+ d\sigma^-  ~ (-1)^a J_{+}^{a} \Omega_{_{ab}} L_{-}^{b}  +\frac{\kappa}{12}\
\int_{_{B_{3}}}d^{3}\sigma ~ (-1)^{a+bc} ~ \varepsilon^{\alpha\beta\gamma}  L_{\alpha}^{a} \Omega_{ad}~ f^d_{~bc} ~ L_{\beta}^{b} L_{\gamma}^{c},~~
\end{eqnarray}
where $J_{+} =(-1)^a  J_{+}^{a} T_{_a} $ is the deformed current which is defined by
\begin{eqnarray}\label{5.2}
J_{\pm} = (1+\omega \eta^{2})\frac{1 \pm \tilde{A} R}{1-\eta^{2}R^{2}} L_{\pm},
\end{eqnarray}
where $\eta, \tilde A$ and  $\kappa$ are three independent real parameters such that the
deformation is measured by means of  $\eta$ and $\tilde A$. The last parameter, $\kappa$, is regarded as the level.
When $\eta= \tilde A=0$ and $\kappa=1$, the action \eqref{5.1} is nothing but that of the undeformed WZW model.
The operator $R$ in \eqref{5.2} is a linear map
from the Lie superalgebra $\G$ to itself, $R: {\G} \rightarrow {\G}$. It is a
super skew-symmetric solution of the (m)GCYBE on ${\G}$. That is to say, for any $X, Y \in \G$ it satisfies
\begin{eqnarray}\label{5.3}
[R(X),R(Y)]-R\big([R(X),Y]+[X,R(Y)]\big)=\omega [X,Y],
\end{eqnarray}
where $\omega$ is a constant parameter which can be normalized by rescaling $R$. The above equation can be generalized to
the mGCYBE if one sets $\omega=\pm1$, while the case $\omega=0 $ gives us the homogeneous GCYBE. It is also worth noting that
the super skew-symmetric condition of the linear $R$-operator requires
\begin{eqnarray}\label{5.4}
\big<R(X) , Y\big>+ \big<X  , R(Y)\big> =0.
\end{eqnarray}
The linear operator $R$ is associated to a classical $r$-matrix\footnote{Hereafter,
we will refer to that as $r$-matrix for simplicity.} which has an important role in the deformation process.
The relationship between them is given by the following formula
\begin{eqnarray}\label{5.5}
R(X)=\big<r ~, ~1 \otimes X\big>,
\end{eqnarray}
for any $X \in \G$, where the inner product is evaluated on the second site of the
r-matrix. Given Lie superalgebra $\G$ with the basis $\{T_{_a}\}$ one may define an $r$-matrix
$r \in {\G} \otimes {\G}$\footnote{Note that the r-matrix is a solution of the following standard (m)GCYBE \cite{N.A,J.z,er.crm}
\begin{eqnarray}\label{5.6}
[[r , r]]\equiv [r_{_{12}} , r_{_{13}}] +[r_{_{12}} , r_{_{23}}]+[r_{_{13}} , r_{_{23}}] =\omega ~\Omega,
\end{eqnarray}
where $r_{_{12}} =r \otimes 1$, $r_{_{23}} =1 \otimes r$ and $r_{_{13}} = r^{ab}~ T_{_a} \otimes 1 \otimes T_{_b}$;
moreover, $\Omega \in \Lambda^3 (\G)$ is the canonical triple tensor Casimir of G.
Notice that the standard form of the (m)GCYBE is equivalent to \eqref{5.3}.} in the form
$r = r^{ab}~ T_{_a} \otimes T_{_b}$, where a sum over repeated indices is implied.
When the $r$-matrix is a super skew-symmetric solution of \eqref{5.6}, namely, $r^{ab} =- (-1)^{ab}~ r^{ba}$, then we can rewrite the $r$-matrix in the form of
\begin{eqnarray}\label{5.7}
r=\frac{1}{2} r^{ab} ~T_{_a} \wedge T_{_b},
\end{eqnarray}
where wedge denotes a graded anti-symmetric tensor product, i.e., $T_{_a} \wedge T_{_b}=T_{_a} \otimes T_{_b} -(-1)^{ab}~ T_{_b} \otimes T_{_a}$.
Notice that the $r$-matrix is considered to be even as $r \in \G_{_B} \wedge \G_{_B} \oplus \G_{_F} \wedge  \G_{_F}$
such that $r^{ab} =0$ if $|a| \neq |b|$.
In other words, fermions with bosons cannot be mixed (grading is preserved).
Accordingly, the r-matrix can be written into the form
\begin{eqnarray}\label{5.8}
r= {r}_{_B}^{ij}  ~ K_i \otimes K_j + {r}_{_F}^{\alpha \beta} ~ S_\alpha  \otimes S_\beta,
\end{eqnarray}
where $\{K_{_i}\}_{i=1}^{m}$ and $\{S_{_\alpha}\}_{\alpha=m+1}^{m+n}$ are the respective bosonic and fermionic basis of
a Lie superalgebra $\G=\G_{_B} \oplus  \G_{_F}$ of superdimension $(m|n)$.

Making use of the fact that in $r^{ab}$, $|a|+|b|=0$, and
expanding $X$ and $R$ in terms of the bases of $\G$ as $X= (-1)^{a}~ x^{a} T_{_a}$ and $R= (-1)^{b}~ R_a^{~b} T_{_b}$, and then substituting
\eqref{5.7} into \eqref{5.5} one obtains that
\begin{eqnarray}\label{5.9}
R_a^{~b}=-(-1)^{ac} ~\Omega_{ac}  ~  r^{cb}.
\end{eqnarray}
Matrices such as $\Omega_{ab}$ and $R_a^{~b}$ are also considered similar to $r^{ab}$, that is,
one considers for them $|a|+|b|=0$.
Accordingly, the (m)GCYBE \eqref{5.3} can be rewritten into the following form:
\begin{eqnarray}\label{5.10}
(-1)^k~R_a^{~c}~f^k_{~cd} R_b^{~d}-(-1)^b~R_a^{~c}~f^d_{~cb} R_d^{~k}-(-1)^a~R_b^{~c}~f^d_{~ac} R_d^{~k} = \omega  ~ f^k_{~ab}.
\end{eqnarray}
It would also be useful to obtain the matrix form of the above equations. Using the matrix representation of the structure constants,
$f^c_{~ab} =- ({\cal Y}^c)_{ ab}$ one obtains
\begin{eqnarray}\label{5.11}
(-1)^d~R~{\cal Y}^k R^{^{st}}-(-1)^c~R ({\cal Y}^d R_d^{~k})- ({\cal Y}^d R_d^{~k}) R^{^{st}} =(-1)^k~ \omega {\cal Y}^k,
\end{eqnarray}
where index $d$ in the first term of the left hand side denotes the column of matrix ${\cal Y}^k$,
while in the second term, $c$ corresponds to the row of matrix ${\cal Y}^d$.
In the next section, we employ the above formulation to obtain the linear $R$-operators and $r$-matrices for the $({\C}_0^5 + \A)$
Lie superalgebra similar to what was done for the $gl(1|1)$ and $({\C}^3 +{\A})$ Lie superalgebras \cite{Epr2}.
Using the obtained $R$-operators we will find YB deformation of WZW model based on the $(C_0^5 + A)$.

\section{YB deformation of WZW model on the $(C_0^5 + A)$}
\label{Sec.6}

In this section we first solve the (m)GCYBE \eqref{5.11} in order to obtain the
$R$-operators and inequivalent r-matrices for the $({\C}_0^5 + {\A})$.
The resulting $R$-operators help us to construct the YB deformation of the $(C_0^5 + A)$ WZW model.
In section \ref{Sec.2}, we constructed the WZW model based on the $(C_0^5 + A)$ by considering
an element of the supergroup as in \eqref{2.18}.
Let us build the WZW model based on the $(C_0^5 + A)$ by choosing an element of the supergroup as in \eqref{2.8}.

\subsection{ WZW model on the $(C_0^5 + A)$ by parametrization \eqref{2.8}}
Here we parametrize an element of the $(C_0^5 + A)$ supergroup
with the coordinates $(y, x; \psi,  \chi)$ so that its elements can be written as in \eqref{2.8}, i.e.,
\begin{eqnarray}
g = e^{\chi T_4}~e^{y T_1}~e^{x T_2}~e^{\psi T_3}. \label{6.1}
\end{eqnarray}
Then, one calculates
\begin{eqnarray}\label{6.2}
g^{-1} \partial_{_{\pm}} g &=& \partial_{_{\pm}} y ~T_{_{1}} + \Big[\partial_{_{\pm}} x + \partial_{_{\pm}} \chi (-\frac{\chi}{2} + \psi \sin y)- \partial_{_{\pm}} \psi \frac{\psi}{2}\Big] T_{_{2}} \nonumber\\
&&~~+\big(\partial_{_{\pm}} \psi - \partial_{_{\pm}} \chi~ \sin y\big) T_{_{3}}+ \big(-\partial_{_{\pm}} y~ \psi + \partial_{_{\pm}} \chi ~\cos y\big) T_{_{4}},
\end{eqnarray}
for which the $L^{\hspace{-0.5mm}a}_{\alpha}$'s are obtained to be of the following form
\begin{eqnarray}\label{6.3}
L^{\hspace{-0.5mm}1}_{\pm}&=&\partial_{_{\pm}} y,\nonumber\\
L^{\hspace{-0.5mm}2}_{\pm}&=&\partial_{_{\pm}} x + \partial_{_{\pm}} \chi (-\frac{\chi}{2} + \psi \sin y)- \partial_{_{\pm}} \psi \frac{\psi}{2},\nonumber\\
L^{\hspace{-0.5mm}3}_{\pm}&=& - \partial_{_{\pm}} \psi+ \partial_{_{\pm}} \chi~ \sin y,\nonumber\\
L^{\hspace{-0.5mm}4}_{\pm}&=& \partial_{_{\pm}} y~ \psi - \partial_{_{\pm}} \chi ~\cos y.
\end{eqnarray}
Analogously, the ad-invariant metric on the $({\C}_0^5 + {\A})$ is as in \eqref{2.7}.
Using these and some algebraic calculations, the WZW action on the $({C}_0^5 + {A})$ are worked out
\begin{eqnarray}\label{6.4}
S_{_{_{\hspace{0mm}WZW}}}(g) &=& \frac{1}{2} \int d \sigma^+ d \sigma^-\Big[\beta\partial_{_{+}} y \partial_{_{-}} y+\partial_{_{+}} y \partial_{_{-}} x+ \partial_{_{+}} x \partial_{_{-}} y
+\partial_{_{+}} y  \chi  \partial_{_{-}} \chi\nonumber\\
&&~~~~~~~~~~~~~~~~~~~~~~~~~~~~~~~~~~~~~~~+\partial_{_{+}} \psi  \psi  \partial_{_{-}} y -2\partial_{_{+}} \psi \cos y~ \partial_{_{-}} \chi\Big].
\end{eqnarray}
By regarding this action as a $\sigma$-model action of the form
\eqref{2.4}, we can read off the line element and $B$-field as follows:
\begin{eqnarray}
ds^2 &=&  \beta dy^2+ 2dydx - \psi ~ dy d\psi + \chi dy d\chi -2 \cos y~ d\psi d\chi,~~~\label{6.5}\\
B &=& \frac{\psi}{2} ~dy \wedge d\psi + \frac{\chi}{2} ~dy \wedge d\chi -\cos y~
d \psi \wedge d \chi.\label{6.6}
\end{eqnarray}
The metric \eqref{6.5} is flat in the sense that its scalar curvature vanish.
One quickly finds that the only non-zero component of Ricci tensor
is ${\cal R}_{_{yy}}=-{1}/{2}$; and as the only non-zero components of $B$-filed are $B_{_{y \psi}}=\psi/2,
B_{_{y \chi}}=\chi/2$ and $B_{_{\psi \chi}}=\cos y$, the only
non-zero components of the field strength $H$, which are obtained from formula \eqref{2.5}, are
$H_{_{y \psi \psi}}=H_{_{y \chi \chi}} =1,$ and $ H_{_{y \psi \chi}}= -  \sin y$.
Putting these pieces together, one verifies equations \eqref{2.24} with $\Lambda=0$ and
$\Phi =c_{_0} y +c_{_1}$ for some constants $c_{_0}$ and $c_{_1}$.
Hence, as expected, the conformal invariance of the model is guaranteed up to the one-loop order.

\subsection{\label{Sec.6.1} $R$-operators and r-matrices of the $({\C}_0^5 + {\A})$}

So far, the r-matrices corresponding to the $({\C}_0^5 + {\A})$ have not been calculated.
Here we obtain the corresponding r-matrices as the solutions of (m)CYBE and show that they are split into five inequivalent classes.
Before proceeding to solve the (m)GCYBE \eqref{5.11}, let us assume that
the most general super skew-symmetric r-matrix $r \in {\cal G}_{_{(2|2)}} \otimes {\cal G}_{_{(2|2)}}$ has the following form:
\begin{eqnarray}\label{6.7}
r= r^{ab} T_{_a} \otimes T_{_b} =  m_1 T_{_1} \wedge T_{_2} +m_2 T_{_3} \wedge T_{_4}
+\frac{1}{2} m_3 T_{_3} \wedge T_{_3}+\frac{1}{2} m_4 T_{_4} \wedge T_{_4},
\end{eqnarray}
where $m_i$ are some real parameters.
Inserting \eqref{6.7} and \eqref{2.7} into \eqref{5.9} one can obtain the general form of the corresponding $R$-operator, giving us
\begin{eqnarray}\label{6.8}
{R}_{_a}^{~b}=\left( \begin{tabular}{cccc}
                $m_1$ & -$\beta m_1$  &  0  & 0\\
                0     & -$m_1$  &  0        & 0\\
                0     & 0  &  $m_2$  & $m_4$\\
                0     & 0  & -$m_3$ & -$m_2$\\
                 \end{tabular} \right).
\end{eqnarray}
In order to solve equation \eqref{5.11} for the $({\C}_0^5 + {\A})$
we need to the matrix representation of the structure constants given by \eqref{2.17}.
Then, by substituting $R$-operator of \eqref{6.8} into equation \eqref{5.11},
the general solution of the (m)GCYBE is split into two families ${R_{_I}}_{_a}^{~b}$ and ${R_{_{II}}}_{_a}^{~b}$
such that the solutions are, in terms of the constants
$\beta, \omega$,  $m_1$, $m_3$, given by
{\small \begin{eqnarray}
{R_{_I}}_{_a}^{~b}=\left( \begin{tabular}{cccc}
                0  & 0  &  0  & 0\\
                0  & 0  &  0  & 0\\
                0  & 0  & $\pm \sqrt{\omega - m_3^2}$  & -$m_3$\\
                0  & 0  & -$m_3$ & $\mp \sqrt{\omega - m_3^2}$\\
                 \end{tabular} \right),~~~
                 {R_{_{II}}}_{_a}^{~b}=\left( \begin{tabular}{cccc}
                $m_1$ & -$\beta m_1$  &  0  & 0\\
                0     & -$m_1$  &  0        & 0\\
                0     & 0  &  0  & $\pm \sqrt{\omega}$\\
                0     & 0  & $\mp \sqrt{\omega}$ & 0\\
                 \end{tabular} \right).\label{6.9}
\end{eqnarray}}
The $r$-matrices corresponding to the above $R$-operators can be obtained by applying equations \eqref{2.7} and \eqref{5.9}.
They are then read
\begin{eqnarray}
{r_{_{I}}}&=&\pm \sqrt{\omega - m_3^2} ~T_{_3}  \wedge T_{_4} +\frac{m_{_3}}{2} \big(T_{_3}  \wedge T_{_3} -T_{_4}  \wedge T_{_4}\big),\label{6.10}\\
{r_{_{II}}}&=& m_{_1} ~T_{_1} \wedge T_{_2} \pm \frac{\sqrt{\omega}}{2} \big(T_{_3}  \wedge T_{_3} +T_{_4}  \wedge T_{_4}\big).\label{6.11}
\end{eqnarray}
The next step is that to specify the exact value of the parameters $m_i$ of the solutions above,
in such a way that one determines the inequivalent r-matrices for the $({\C}_0^5 + {\A})$. In this regard, there is a Proposition \cite{{Epr1},{Epr2}}
stating that two r-matrices $r$ and $r'$ of a Lie superalgebra $\G$ are equivalent if one can be obtained from the other by means of
a change of basis which is an automorphism $A$ of $\G$, such that
\begin{eqnarray}\label{6.12}
r^{ab} = (-1)^d~ (A^{^{st}})^a_{~c}~ {r^\prime}^{cd}~{A_d}^b.
\end{eqnarray}
According to formula \eqref{6.12} one must find the automorphism supergroup of Lie superalgebra $\G$
which preserves (a) the parity of the generators (they cannot mix fermions with bosons), and (b) the structure constants $f^c_{~ab}$.
Therefore it is crucial for our further considerations to identify the
supergroup of automorphisms of the $({\C}_0^5 + {\A})$.
We define the action of the automorphism $A$ on $\G$ by the transformation $T'_a = (-1)^b~ {A}_a^{~b}~ T_b$.
The set of automorphisms of $({\C}_0^5 + {\A})$ is generated by two transformations:
\begin{eqnarray}
T'_{_1} = T_{_1}+c T_{_2},~~~~~T'_{_2} = (a^2+b^2) T_{_2},~~~~~~T'_{_3} = -a T_{_3} +b T_{_4},~~~~~~T'_{_4} = -b T_{_3}-a T_{_4},~~~~\label{6.13}
\end{eqnarray}
and
\begin{eqnarray}
T'_{_1} = -T_{_1}+c T_{_2},~~~~~T'_{_2} = (a^2+b^2) T_{_2},~~~~~~T'_{_3} = a T_{_3} -b T_{_4},~~~~~~T'_{_4} = -b T_{_3}-a T_{_4},~~~~\label{6.14}
\end{eqnarray}
where $a, b, c$ are some arbitrary real constants.
The bases $\{T'_a\}$ obey the same (anti-)commutation relations as $\{T_a\}$.
When taken into account, the above transformations lead to a conclusion that the parameters $m_{_1}$ and $m_{_3}$
in \eqref{6.10} and \eqref{6.11} can be scaled out to take the value of $0$ or $1$.
Now, by using the transformations \eqref{6.13} and \eqref{6.14} and by employing formula \eqref{6.12}
we arrive at five families of inequivalent r-matrices for the $({\C}_0^5 + {\A})$ whose representatives can be described by means of the following Theorem.
\begin{theorem}\label{thm1.A}
Any r-matrix of the $({\C}_0^5 + {\A})$ Lie superalgebra as a solution of the (m)GCYBE belongs
just to one of the following five inequivalent classes
\begin{eqnarray*}
{r_{_i}}&=& T_{_1} \wedge T_{_2},~~~~~~~~~~~~~~~~~~~\nonumber\\
{r_{_{ii}}}&=&\frac{1}{2} \big(T_{_3}  \wedge T_{_3} +T_{_4}  \wedge T_{_4}\big),\nonumber\\~~~~~~~~~~~~~~~~~~~
{r_{_{iii}}}&=&-\frac{1}{2} \big(T_{_3}  \wedge T_{_3} +T_{_4}  \wedge T_{_4}\big),\nonumber\\~~~~~~~~~~~~~~~~~~~
{r_{_{iv}}}&=& T_{_1} \wedge T_{_2} + \frac{p}{2} \big(T_{_3}  \wedge T_{_3} +T_{_4}  \wedge T_{_4}\big);~~~p \neq 0,\nonumber\\
{r_{_{v}}}&=& T_{_3}  \wedge T_{_4}.\nonumber~~~~~~~~~~~~~~~~~~~
\end{eqnarray*}
\end{theorem}
Note that the $r$-matrix ${r_{_i}}$ satisfies the (not modified) GCYBE, whereas the rest of the $r$-matrices do not.
Therefore, the ${r_{_i}}$ is a homogenous solution.
Indeed, the $r$-matrices ${r_{_{ii}}}$, ${r_{_{iii}}}$ and ${r_{_{v}}}$ satisfy mGCYBE with $\omega=1$, while
${r_{_{iv}}}$ with $\omega=p^2$.
The parameter $p$ is present in ${r_{_{iv}}}$ as it cannot be removed by means of the automorphism transformations \eqref{6.13} and \eqref{6.14}.

Before closing this subsection, let us look at the Abelian and unimodularity conditions on the $r$-matrices of the $({\C}_0^5 + {\A})$.
As mentioned in \cite{K.Yoshida}, the target spacetime of YB deformations of $AdS_5 \times S^5$ based on the homogeneous CYBE \cite{{Delduc2},{Delduc3}}
satisfies the equations of motion of type IIB supergravity if
the r-matrix satisfies the unimodularity condition \cite{Wulff1}. If not, the background is a
solution of so-called generalized type IIB supergravity \cite{{Arutyunov2},{Wulff2}}.
So, non-unimodular YB deformations result in solutions of generalized supergravity.
In turn, the r-matrices may be crudely divided into two families:
Abelian and non-Abelian. It has been proved that Abelian r-matrices correspond to
TsT transformations \cite{Osten}, thus ensuring that the corresponding YB deformation is a solution of the supergravity.
For non-Abelian r-matrices, a further unimodularity condition on the
r-matrix \cite{Wulff1} distinguishes valid supergravity backgrounds from solutions to generalized supergravity.
The r-matrix is called Abelian if $[T_a , T_b] =0$ and unimodular if it satisfies the following condition
\begin{eqnarray}
r^{ab}~ [T_a , T_b] =0.\label{6.15}
\end{eqnarray}
Using \eqref{6.15} together with \eqref{2.17} we find that the
only the $r$-matrices ${r_{_i}}$ and ${r_{_v}}$ are Abelian and also unimodular,
while the rest denote the non-Abelian and non-unimodular r-matrices.

\subsection{\label{Sec.6.2} YB deformed backgrounds of the $({C}_0^5 + {A})$ WZW model}

Let us turn into the  main goal of this section which is nothing but calculating the
YB deformations of the $({C}_0^5 + {A})$ WZW model.
Having $R$-operators one can calculate the deformed currents.
Now we use formulas \eqref{2.7} and \eqref{5.9}
to obtain all $R$-operators corresponding to the inequivalent r-matrices of Theorem 6.1.
Then we employ equation \eqref{5.2} to obtain the deformed currents $J_{\pm}$.
To this end, one may write down \eqref{5.2} as follows \cite{Epr2}:
\begin{eqnarray}\label{6.16}
J^a_{\pm} -(-1)^{b+c}~\eta^{2} J^b_{\pm} ~R_b^{~c}~R_c^{~a}= (1+\omega \eta^{2})\big[L^{a}_{_\pm} \pm (-1)^{b}~ \tilde{A} L^{b}_{_\pm} ~R_b^{~a}\big].
\end{eqnarray}
Finally by using the resulting $R$-operators and also by utilizing relations
\eqref{6.16} and \eqref{6.3} together with the action \eqref{5.1} one can obtain all YB deformed backgrounds of the $({C}_0^5 + {A})$ WZW model.
Below, we obtain the deformed backgrounds including metric and $B$-field, and then check the conformal invariance conditions of the deformed models up to the one-loop order.

\subsubsection{\label{Sec.6.2.1} Deformation with the ${r_{_{v}}}$}
Let us now present the YB deformed background whose initial input is the $r$-matrix ${r_{_v}}$.
The corresponding $R$-operator can be obtained by using formulas \eqref{2.7} and \eqref{5.9}.
Then, one can employ formula \eqref{6.16} to get the corresponding deformed currents, giving
\begin{eqnarray}\label{6.17}
J^1_{\pm} &=& (1+\eta^2)L^{1}_{_\pm},~~~~~~~~~~~~~~~~~~~~~J^2_{\pm} = (1+\eta^2)L^{2}_{_\pm},\nonumber\\
J^3_{\pm} &=& \frac{(1+\eta^2)(1-{\tilde A})}{1-\eta^2}  L^{3}_{_\pm},~~~~~~~~~~~J^4_{\pm} = \frac{(1+\eta^2)(1+{\tilde A})}{1-\eta^2}  L^{4}_{_\pm}.
\end{eqnarray}
To obtain the explicit form of $J_{\pm}$'s one must use \eqref{6.3}. Then, by applying the action \eqref{5.1}, the deformed background including line element and $B$-field are, respectively, given by	
\begin{eqnarray}
ds_{_{def}}^2 &=& (1+\eta^2)\Big[\beta dy^2+ 2dydx -\frac{1+\eta^2}{1-\eta^2} \psi ~ dy d\psi + \big(\chi + \frac{2\eta^2}{1-\eta^2} \psi \sin y\big)  dy d\chi \nonumber\\
&&~~~~~~~~~~~~~~~~~~~~~~~~~~~~~~~~~~~~~~~~~~~~~~~~~~~~~~~~-\frac{2}{1-\eta^2} \cos y~ d\psi d\chi\Big],\label{6.18}\\
B_{_{def}} &=& \big(\frac{\kappa}{2} - \frac{{\tilde A} (1+\eta^2)}{1-\eta^2}\big) {\psi} dy \wedge d\psi  -
\big(\kappa - \frac{2{\tilde A} (1+\eta^2)}{1-\eta^2}\big)\cos y~
d \psi \wedge d \chi\nonumber\\
&&~~~~~~~~~~~~~~~~~~~~~~~~~~~+ \frac{\kappa \chi}{2} ~dy \wedge d\chi -\frac{{\tilde A} (1+\eta^2)}{2(1-\eta^2)} \sin (2y)~
d \chi \wedge d \chi.\label{6.19}
\end{eqnarray}
Thus we have built a new integrable $\sigma$-model whose background is described by the metric \eqref{6.18} and $B$-filed \eqref{6.19}.

As we know, YB deformed WZW models are integrable \cite{Delduc4}. Now one question arises
whether the WZW model remains conformally invariant after the deformation (at least up to the one-loop order perturbatively)?
In order to answer this question on has to investigate the conformal invariance conditions (the beta
functions) for the background of the deformed models\footnote{Note that the one-loop beta function equations on supermanifolds (in Dewitt's notation) were written in \cite{ER5}.
So far, at higher loops the equations on supermanifolds have not been written.
Accordingly, we investigate the conformal invariance conditions of YB deformed backgrounds only up to the one-loop order.}.
Accordingly, it seems  to  be  of  interest  to check the conformal invariance conditions, equations \eqref{2.24}, for the background
defined by the metric \eqref{6.18} and $B$-filed \eqref{6.19}.
To this end, one quickly finds that only the non-zero component of Ricci tensor
is ${\cal R}_{_{yy}}=-\frac{1}{2} (1-\eta^2)^2$, and thus the scalar curvature vanishes.
Furthermore, the non-zero components of field strength corresponding to $B$-filed \eqref{6.19} are obtained to be
\begin{eqnarray}\label{6.20}
H_{_{y \psi \psi}}=\kappa+ \frac{2 {\tilde A} (1+\eta^2)}{\eta^2-1}, ~~
H_{_{y \psi \chi}}= - H_{_{y \psi \psi}} \sin y,~~~H_{_{y \chi \chi}}=\kappa- \frac{2 {\tilde A} (1+\eta^2)}{\eta^2-1} \cos (2y).
\end{eqnarray}
Using these, equations \eqref{2.24} with $\Lambda=0$ and the dilaton field
$\Phi =c_{_0} y +c_{_1}$, for some constants $c_{_0}, c_{_1}$, are satisfied
if the following relation holds between the constants $\kappa, \eta$ and ${\tilde A}$:
\begin{eqnarray}\label{6.21}
\kappa = \frac{1+\eta^2}{1-\eta^2} \sqrt{4 {\tilde A}^2 +(1-\eta^2)^2}.
\end{eqnarray}

Let us further highlight how the YB deformed background (\eqref{6.18} and \eqref{6.19}) differs from the undeformed one (\eqref{6.5} and \eqref{6.6}).
As is evident, the deformation has affected the coupling of the fourth term of the metric, as well as the last term of the $B$-field.
As shown above, the deformation creates a new geometry with a different Ricci curvature, as well as shifting the $B$-field.
Most interestingly, it can be shown that the isometric symmetries of the metric change by the deformation. In order to investigate this,
one must apply the graded Killing equation \eqref{1}.
First, for the undeformed metric given by equation \eqref{6.5}, we find that the metric admits four bosonic Killing vectors
\begin{align}\label{55}
K^B_{_1} &= \overrightarrow{\frac{\partial}{\partial x}},~~~~~~~~~~~~~~~~~~~~~~~~~~
K^B_{_2} = \sec y (1+\sin y)  (\psi-\chi) \big(\overrightarrow{\frac{\partial}{\partial \psi}} + \overrightarrow{\frac{\partial}{\partial \chi}}\big),\nonumber\\
K^B_{_3} &=\frac{\cos y}{(1+\sin y)}  (\psi+\chi) \big(-\overrightarrow{\frac{\partial}{\partial \psi}} + \overrightarrow{\frac{\partial}{\partial \chi}}\big),~~~~~~~~~~~
K^B_{_4} = \frac{1}{2} \big(\chi \overrightarrow{\frac{\partial}{\partial \psi}} + \psi  \overrightarrow{\frac{\partial}{\partial \chi}}\big),
\end{align}
together with the following four fermionic Killing vectors
\begin{align}\label{66}
K^F_{_1} &= \frac{\psi}{2} \overrightarrow{\frac{\partial}{\partial x}} +  \overrightarrow{\frac{\partial}{\partial \psi}},~~~~~~~~~~~~~~~~~~
K^F_{_2} = -\frac{\chi}{2} \overrightarrow{\frac{\partial}{\partial x}} +  \overrightarrow{\frac{\partial}{\partial \chi}},\nonumber\\
K^F_{_3} &=\frac{1}{2} (-\psi \sec y+\chi \tan y) \overrightarrow{\frac{\partial}{\partial x}} + \sec y \overrightarrow{\frac{\partial}{\partial \psi}} + \tan y \overrightarrow{\frac{\partial}{\partial \chi}},\nonumber\\
K^F_{_4} &= \frac{1}{2} (-\psi \tan y+\chi \sec y) \overrightarrow{\frac{\partial}{\partial x}} +\tan y  \overrightarrow{\frac{\partial}{\partial \psi}} + \sec y \overrightarrow{\frac{\partial}{\partial \chi}},
\end{align}
while the deformed metric \eqref{6.18} admits only one bosonic Killing vector $K^B_{_1}=4 \overrightarrow{\frac{\partial}{\partial x}}$, and
the following two fermionic Killing vectors
\begin{align}\label{77}
K^F_{_1} &=(-\psi \sec y +\chi \tan y) \overrightarrow{\frac{\partial}{\partial x}} +  2 \sec y  \overrightarrow{\frac{\partial}{\partial \psi}}+ 2 \tan y \overrightarrow{\frac{\partial}{\partial \chi}},\nonumber\\
K^F_{_2} &= \frac{1}{2} \sec y  \big[\chi \cos(y \eta^2)-\psi \sin (y+ y \eta^2)\big]  \overrightarrow{\frac{\partial}{\partial x}}\nonumber\\
&~~~~~~~~~~~~~~~~~~~~+ \sec y \sin (y- y \eta^2) \overrightarrow{\frac{\partial}{\partial \psi}} + \sec y \cos(y \eta^2) \overrightarrow{\frac{\partial}{\partial \chi}}.
\end{align}
It can be investigated that these supervectors span the $(1|2)$-dimensional Lie superalgebra $(\A_{1,1}+2 \A)^2$ \cite{B} with
 the following non-zero Lie superbrackets
\begin{align}\label{88}
\{K^F_{_1}  , K^F_{_1}\} = - K^B_{_1},~~~~~~~~~~~~\{K^F_{_2} , K^F_{_2}\} = K^B_{_1}.
\end{align}
The effect of deformation is clearly seen in the supervector $K^F_{_2}$.
This indicates that the isometric symmetry of the metric is broken by the deformation.

\subsubsection{\label{Sec.6.2.2} Deformation with the ${r_{_{\epsilon_{_1},\epsilon_{_2}}}}$}

Here we consider the $r$-matrices ${r_{_i}}$, ${r_{_{ii}}}$, ${r_{_{iii}}}$ and ${r_{_{iv}}}$ in the form of a $r$-matrix
as follows:
\begin{eqnarray}\label{6.22}
{r_{_{\epsilon_{_1},\epsilon_{_2}}}} =\epsilon_{_1} T_{_1} \wedge T_{_2} + \frac{\epsilon_{_2}}{2} \big(T_{_3}  \wedge T_{_3} +T_{_4}  \wedge T_{_4}\big)=
\begin{cases}
{r_{_{i}}}  ~& \epsilon_{_1}=1, \epsilon_{_2}=0,\\
{r_{_{ii}}} ~& \epsilon_{_1}=0, \epsilon_{_2}=1,\\
{r_{_{iii}}} ~& \epsilon_{_1}=0, \epsilon_{_2}=-1,\\
{r_{_{iv}}} ~& \epsilon_{_1}=1, \epsilon_{_2}=p.
\end{cases}
\end{eqnarray}
For this $r$-matrix, one can get the corresponding $R$-operator by using formulas \eqref{2.7} and \eqref{5.9}, giving us
\begin{eqnarray}\label{6.23}
{R}_{_a}^{~b}=\left( \begin{tabular}{cccc}
               $\epsilon_{_1}$  & -$\beta \epsilon_{_1}$  &  0  & 0\\
                0  & -$\epsilon_{_1}$ &  0  & 0\\
                0  & 0  & 0  & $\epsilon_{_2}$\\
                0  & 0  & -$\epsilon_{_2}$ & 0\\
                 \end{tabular} \right).
\end{eqnarray}
Then, using \eqref{6.16} we find that
\begin{eqnarray}\label{6.24}
J^1_{\pm} &=& \frac{(1+\omega \eta^2)(1+\tilde A \epsilon_{_1})}{1- \epsilon_{_1}\eta^2}  L^{1}_{_\pm},~~~~~~~~~~
J^2_{\pm} = \frac{1+\omega \eta^2}{1- \epsilon_{_1}\eta^2}
\big[(1-\tilde A \epsilon_{_1})L^{2}_{_\pm} - \beta \tilde A \epsilon_{_1} L^{1}_{_\pm} \big],\nonumber\\
J^3_{\pm} &=&\frac{1+\omega \eta^2}{1+ {\epsilon^2_{_2}} \eta^2}
\big[L^{3}_{_\pm} +  \tilde A \epsilon_{_2} L^{4}_{_\pm} \big],~~~~~~~~~
J^4_{\pm} = \frac{1+\omega \eta^2}{1+ {\epsilon^2_{_2}}\eta^2}
\big[L^{4}_{_\pm} - \tilde A \epsilon_{_2} L^{3}_{_\pm} \big].
\end{eqnarray}
Finally, inserting \eqref{6.3} into \eqref{6.24} and then applying \eqref{5.1}, the deformed background reads	
\begin{eqnarray}
ds_{_{def}}^2 &=& (1+\omega \eta^2)\Big[\frac{1}{1-\epsilon_{_1} \eta^2} (\beta dy^2+ 2dydx) +\big(\frac{1}{1-\epsilon_{_1} \eta^2}  - \frac{2}{1+{\epsilon_{_2}}^2 \eta^2}\big) \psi ~ dy d\psi \nonumber\\
&&~~~~~~~~~~~~~~+ \Big(\frac{1}{1-\epsilon_{_1} \eta^2}\chi + 2\big(\frac{1}{1+{\epsilon_{_2}}^2 \eta^2}-  \frac{1}{1-\epsilon_{_1} \eta^2}\big) \psi \sin y\Big)  dy d\chi \nonumber\\
&&~~~~~~~~~~~~~~-\frac{2}{1+{\epsilon_{_2}}^2 \eta^2} \cos y~ d\psi d\chi\Big],\label{6.25}\\
B_{_{def}} &=& \frac{1}{2}\Big(\kappa + \frac{\tilde A \epsilon_{_1}(1+\omega \eta^2)}{1-\epsilon_{_1} \eta^2}\Big) {\psi} dy \wedge d\psi  +\frac{1}{2}
\Big[\kappa  \chi+ \frac{\tilde A \epsilon_{_1}(1+\omega \eta^2)}{1-\epsilon_{_1} \eta^2} (\chi -2 \psi \sin y)\Big]
dy \wedge d\chi~~\nonumber\\
&&~~- \kappa  \cos y ~d\psi \wedge d\chi.\label{6.26}
\end{eqnarray}
The metric is flat in the sense that its scalar curvature vanishes. We furthermore find that only the non-zero component of Ricci tensor
is ${\cal R}_{_{yy}}=-{(1+{\epsilon_{_2}}^2\eta^2)^2}/{2 (1-{\epsilon_{_1}} \eta^2)^2}$
which shows that the deformation has created a new geometry with a different Ricci curvature.
Notice that we have ignored the terms $\frac{\tilde A \epsilon_{_1}(1+\omega \eta^2)}{1-\epsilon_{_1} \eta^2} dy \wedge dx$
and $-\frac{\tilde A \epsilon_{_2}(1+\omega \eta^2)}{2(1+ {\epsilon^2_{_2}} \eta^2)} \big(d\psi \wedge d\psi + d\chi \wedge d\chi\big)$
that appeared in the $B$-field part as total derivative terms. Hence, we find that the non-zero components of field strength are
\begin{eqnarray}
H_{_{y \psi \psi}}=\kappa+ \frac{{\tilde A} \epsilon_{_1} (1+\omega \eta^2)}{1-\epsilon_{_1} \eta^2}, ~~~~~H_{_{y \chi \chi}}=  H_{_{y \psi \psi}},~~~~~
H_{_{y \psi \chi}}= - H_{_{y \psi \psi}} \sin y.
\end{eqnarray}
Putting these pieces together, one can conclude that the deformed background above with zero cosmological constant, $\Lambda=0$, and
dilaton filed $\Phi =c_{_0} y +c_{_1}$ is conformally invariant up to the one-loop order provided that
\begin{eqnarray}
\kappa = -\frac{ (1+ {\tilde A} \epsilon_{_1}) (1+\omega \eta^2)}{1-\epsilon_{_1} \eta^2}.
\end{eqnarray}


\section{Summary and concluding remarks}
\label{Sec.7}

In this study, we have investigated different aspects of ungauged and gauged WZW models in superdimension up to $(2|2)$.
The emphasis was on the determination of the geometric data of the corresponding non-linear $\sigma$-models and the discussion of dualities,
specifically from the perspective of super Poisson-Lie symmetry.

First of all, we have reviewed the construction of the WZW models based on the $GL(1|1)$ and $(C^3 + A)$ Lie supergroups, as well as
their super Poisson-Lie symmetry. As a new exact conformal field theory on supergroups,
we have built the WZW model on the $({C}_0^5 + {A})$ Lie supergroup.
Unfortunately, the model did not contain the super Poisson-Lie symmetry.
Then we became interested in constructing gauged WZW models on the supersets GL(1|1)/SO(2), $(C^3 + A)/$SO(2) and  $({C}_0^5 + {A})/$SO(2).
In this regard, we constructed a number of new three-dimensional exact conformal field theories of type $(1|2)$.
Most interestingly, we have shown that the gauged WZW model on the supercoset $(C^3 + A)/$SO(2) has the super Poisson-Lie symmetry,
in such a way that we were able to find the corresponding dual pair. The dual model itself remained conformally invariant up to the one-loop order.

It would be interesting to generalize our results to the cases of
the higher-dimensional Lie supergroups such as OSP$(1|2)$ and OSP$(2|2)$.
This is technically challenging (their bosonic part is non-Abelian, unlike the Lie superalgebras under consideration in the present work)
but should be possible with some guidance from the supergravity solutions that should correspond
to the bosonic version of the models associated with these higher-dimensional supercosets (see \cite{{ref.6},{ref.7}}).
We intend to address this problem in the future.
Recently, the OSP$(1|2)$ Lie supergroup has been of interest from the T-duality perspective \cite{Bielli1} (see, also, \cite{Bielli2}).
Performing the T-dualization of the OSP$(1|2)$ principal chiral model, it has been shown that the super non-Abelian T-dual model does not
satisfy the three-dimensional supergravity constraints, thus falling outside the class
of supergravity backgrounds.

Inspired by a prescription invented by authors of Ref. \cite{Delduc4}, we have found the deformed backgrounds corresponding to the YB deformed WZW model
for the $({C}_0^5 + {A})$ Lie supergroup.
To this end, we obtained the inequivalent solutions of the (m)GCYBE for the $({\C}_0^5 +{\A})$ Lie superalgebra, and showed that the corresponding
$r$-matrices were split into five inequivalent classes.
We found that the only the $r$-matrices ${r_{_i}}$ and ${r_{_v}}$ of Theorem 6.1 were Abelian and also unimodular,
while the rest denoted the non-Abelian and non-unimodular r-matrices.
According to \cite{K.Yoshida}, at the level of string theory, the
condition that the backgrounds of YB deformed models solve the standard supergravity equations of motion
requires the associated classical r-matrices to be unimodular \cite{Wulff1}.
Abelian classical r-matrices are always unimodular, meaning any such classical r-matrix maps a solution of
standard supergravity to a solution of standard supergravity.
Therefore, we expected that the backgrounds deformed by the matrices ${r_{_i}}$ and ${r_{_v}}$ be solutions to the standard supergravity equations,
implying that the YB deformed models are Weyl invariant at the quantum level and thus can be defined a consistent string theory.
In contrast to Abelian ones, non-Abelian classical r-matrices may be non-unimodular,
and indeed the associated backgrounds solve the generalized supergravity equations \cite{Arutyunov2}, but not the standard ones.
According to the above statement, the YB deformed backgrounds associated to non-unimodular r-matrices ${r_{_{ii}}}, {r_{_{iii}}}$ and ${r_{_{iv}}}$ can be solutions to the generalized equations.
Recently, we have written the generalized supergravity equations introduced by Arutyunov {\it et al.} on a supermanifold $\M$, and then
called them the graded  generalized supergravity equations \cite{EYR}.
Note that the  generalized equations on supermanifolds had not been discussed before.
Thus, examining the graded generalized supergravity equations for our YB deformed backgrounds will be a matter of considerable debate.
Of course, it is worth mentioning that the YB deformations of the WZW models based on the $GL(1|1)$ and $(C^3 + A)$ Lie supergroups have already been performed in Ref. \cite{Epr2}.
However, we think that our results are important, in particular in the supergroup case they are rare, much hard technical work was needed to obtain them.
As a future direction, it would be interesting to follow up integrable deformations ($\lambda$-, $\eta$- and YB-types) of the gauged WZW models found in this paper. In this regard, it has recently been shown that \cite{Driezen.Sevrin.Thompson} the integrable deformations of the $\lambda$-type can be constructed for the asymmetrically gauged
WZW models by a modification of the Sfetsos gauging procedure to
account for a possible automorphism that is allowed in $G/G$ models.

\subsection*{Declaration of competing interest}

The authors declare that they have no known competing financial
interests or personal relationships that could have appeared to influence
the work reported in this paper.

\subsection*{Acknowledgements}

The authors would like to thank the anonymous referee for invaluable comments and criticisms.
This work has been supported by the research vice chancellor of Azarbaijan Shahid Madani University under research fund No. 1401/537.


\subsection*{Data availability statement}

No data was used for the research described in the article.

\end{document}